\documentclass[journal,romanappendices]{IEEEtran}
\IEEEoverridecommandlockouts

\usepackage{algorithm,algorithmic,amsmath,amssymb,amsthm,bbm,cite,comment,color,graphicx,microtype,setspace,soul,url}
\usepackage[USenglish]{babel}
\usepackage{hyperref}
\usepackage[utf8]{inputenc}
\usepackage[T1]{fontenc}
\usepackage{subcaption}
\usepackage{multirow}

\usepackage[figurename=Fig., labelsep=period, font=small]{caption}

\usepackage{tikz,pgfplots}
\usetikzlibrary{arrows}
\pgfplotsset{compat=1.17}

\usepackage{titlesec}
\let\subparagraph\relax
\titlespacing{\section}{0pt}{5pt plus 2pt minus 1pt}{3pt plus 1pt minus 1pt} 
\titlespacing{\subsection}{0pt}{4pt plus 2pt minus 1pt}{2pt plus 1pt minus 1pt} 
\usepackage{todonotes}
\usepackage{marginnote}

\usepackage{array}
\usepackage{makecell}

\renewcommand{\a}{\mathbf{a}}

\renewcommand{\d}{\mathbf{d}}
\newcommand{\e}{\mathbf{e}}
\newcommand{\f}{\mathbf{f}}
\newcommand{\g}{\mathbf{g}}
\newcommand{\h}{\mathbf{h}}

\newcommand{\n}{\mathbf{n}}

\newcommand{\p}{\mathbf{p}}
\newcommand{\q}{\mathbf{q}}
\renewcommand{\r}{\mathbf{r}}

\renewcommand{\u}{\mathbf{u}}
\renewcommand{\v}{\mathbf{v}}

\newcommand{\x}{\mathbf{x}}
\newcommand{\y}{\mathbf{y}}
\newcommand{\z}{\mathbf{z}}

\newcommand{\0}{\mathbf{0}}

\newcommand{\A}{\mathbf{A}}
\newcommand{\B}{\mathbf{B}}
\newcommand{\C}{\mathbf{C}}

\newcommand{\G}{\mathbf{G}}
\renewcommand{\H}{\mathbf{H}}
\newcommand{\I}{\mathbf{I}}

\renewcommand{\P}{\mathbf{P}}

\newcommand{\R}{\mathbf{R}}

\newcommand{\V}{\mathbf{V}}

\newcommand{\X}{\mathbf{X}}
\newcommand{\Y}{\mathbf{Y}}
\newcommand{\Z}{\mathbf{Z}}

\newcommand{\zetab}{\boldsymbol{\zeta}}

\newcommand{\mub}{\boldsymbol{\mu}}

\newcommand{\xib}{\boldsymbol{\xi}}

\newcommand{\Sigmab}{\mathbf{\Sigma}}

\newcommand{\Phib}{\mathbf{\Phi}}

\newcommand{\setC}{\mathcal{C}}

\newcommand{\setE}{\mathcal{E}}

\newcommand{\setN}{\mathcal{N}}

\newcommand{\setQ}{\mathcal{Q}}

\newcommand{\setS}{\mathcal{S}}

\newcommand{\Real}{\mbox{$\mathbb{R}$}}
\newcommand{\Compl}{\mbox{$\mathbb{C}$}}

\newcommand{\argmin}{\operatornamewithlimits{argmin}}
\DeclareMathOperator*{\minn}{min}

\newcommand{\blkdiag}{\mathrm{blkdiag}}
\newcommand{\diag}{\mathrm{diag}}

\newcommand{\diff}{\mathrm{d}}
\newcommand{\Exp}{\mathbb{E}}
\newcommand{\herm}{\mathrm{H}}
\renewcommand{\Im}{\mathrm{Im}}

\renewcommand{\Pr}{\mathbb{P}}

\renewcommand{\Re}{\mathrm{Re}}
\newcommand{\sgn}{\mathrm{sgn}}

\newcommand{\tr}{\mathrm{tr}}
\newcommand{\tran}{\mathrm{T}}

\newcommand{\erf}{\mathrm{erf}}

\newtheorem{theorem}{Theorem}

\newtheorem{lemma}{Lemma}
\newtheorem{proposition}{Proposition}

\newcommand{\blue}[1]{\textcolor{blue}{#1}}

\title{Enhanced Uplink Data Detection in Massive MIMO with 1-Bit ADCs: Analysis and Joint Detection}
\author{Amin Radbord, Italo Atzeni, and Antti Tölli
\thanks{The authors are with the Centre for Wireless Communications, University of Oulu, Finland (e-mail: \{amin.radbord, italo.atzeni, antti.tolli\}@oulu.fi).
This work was supported by the Research Council of Finland (336449 Profi6, 348396 HIGH-6G, 357504 EETCAMD, and 369116 6G~Flagship) and by the European Commission (101095759 Hexa-X-II). Part of this work was presented at IEEE ICASSP 2023 \cite{Rad23} and ASILOMAR 2023~\cite{Rad23a}.}}

\begin{document}

\maketitle

\begin{abstract}
We present a new analytical framework for the uplink data detection in massive multiple-input multiple-output systems with 1-bit analog-to-digital converters (ADCs). We first characterize the expected values of the soft-estimated symbols (after the linear receiver and prior to the data detection), which are affected by 1-bit quantization during both the channel estimation and the uplink data transmission. In our analysis, we consider conventional receivers such as maximum ratio combining (MRC), zero forcing, and minimum mean squared error (MMSE), with multiple user equipments (UEs) and correlated Rayleigh fading. Additionally, we design a linear minimum mean dispersion (LMMD) receiver tailored for data detection with 1-bit ADCs, which exploits the expected values of the soft-estimated symbols previously derived. Then, we propose a joint data detection (JD) strategy that exploits the interdependence among the soft-estimated symbols of the interfering UEs, analyzing its symbol error rate (SER), and subsequently introduce a low-complexity variant. These strategies are compared with robust maximum likelihood data detection with 1-bit ADCs. Numerical results examining the SER show that MMSE exhibits considerable performance gains over MRC, whereas the proposed LMMD receiver significantly outperforms the conventional receivers. Lastly, the proposed JD and its low-complexity variant provide a significant boost in comparison with the UE-specific data detection.
\end{abstract}

\vspace{-4mm}


\vspace{-1mm}

\section{Introduction} \label{sec:Intro}

To leverage the wide bandwidths in the sub-THz spectrum, massive multiple-input multiple-output (MIMO) arrays are required at the transmitter and/or receiver to overcome the strong pathloss and penetration loss \cite{Raj20,Atz23,Atz25}. In this respect, fully digital massive MIMO architectures with low-resolution analog-to-digital/digital-to-analog converters (ADCs/DACs) can provide highly flexible beamforming and large-scale spatial multiplexing with modest power consumption and complexity, without excessively compromising the spectral efficiency \cite{Li17,Jac17a,Atz21b}. Remarkably, fully-digital massive MIMO systems with few-bit or even 1-bit ADCs/DACs can outperform hybrid analog-digital ones in terms of both spectral and energy efficiency~\cite{Rot18}.

\subsection{State of the Art} \label{sec:I.A}

Data detection in 1-bit quantized massive MIMO systems has been the subject of many recent studies. 
 
An asymptotic closed-form approximation of the symbol error rate (SER) with the quantized zero-forcing (ZF) precoder was derived in \cite{Sax17} for downlink massive MIMO systems with 1-bit DACs. However, this SER approximation is only applicable when the channel is perfectly known at each user equipment (UE). In the context of uplink data detection, a low-complexity message passing de-quantization (MPDQ) detector was developed in \cite{Wang15}, which exploits both the structured sparsity and the prior probability distribution of the transmitted signal. Assuming perfect channel state information (CSI) and a large number of UEs, an MPDQ algorithm via Gaussian belief propagation based on the Bussgang theorem was presented in \cite{Wat22} for millimeter-wave MIMO systems with low-resolution ADCs. An iterative detection and decoding method with 1-bit ADCs was proposed in \cite{Sha18}, which relies on exchanging soft information in the form of log-likelihood ratio between the detector and the channel decoder.

Generalized approximate message passing (GAMP) and variational approximate message passing (VAMP) algorithms for joint channel estimation and data detection with low-resolution and 1-bit ADCs were developed in \cite{Wen16} and \cite{Zha18}, respectively. Based on numerical results, GAMP outperforms VAMP for sufficiently long pilots and moderate-to-high signal-to-noise ratio (SNR) \cite{Zha18}. However, both methods require several iterations with significant computational complexity per iteration, and their performance degrades severely under imperfect CSI.

A two-stage near maximum likelihood (ML) data detection method with 1-bit ADCs was proposed in \cite{Cho16}, which reduces the number of candidate transmit vectors using the output of the first stage and leverages the structure of the original ML detector to further enhance the performance. Under imperfect CSI, the near ML method in \cite{Cho16} always outperforms GAMP and VAMP in \cite{Wen16} and \cite{Zha18}, respectively. A 1-bit sphere-decoding technique was presented in \cite{Jeo18}, where a list of codewords for sphere decoding is constructed based on the weighted Hamming distance. This technique can be regarded as a low-complexity variant of \cite{Cho16}, although it requires a pre-processing stage with computational complexity that grows exponentially with the number of transmit and receive antennas. A robust ML (RML) method with 1-bit ADCs was proposed in \cite{Ngu21}, which circumvents the non-robustness of conventional ML detection under imperfect CSI by approximating the cumulative distribution function of a Gaussian random variable with a sigmoid function. Gradient descent with step size optimized via deep unfolding is adopted to solve the RML detection problem, and a nearest-neighbor (NN) search is utilized to refine the solution. The method in \cite{Ngu21} was shown to outperform ML (and thus also \cite{Wen16,Zha18,Cho16,Jeo18}) under imperfect CSI. In addition, recent works have shown that machine learning can provide efficient and robust channel estimation and data detection in massive MIMO systems with 1-bit ADCs \cite{Gao17,Ngu21,Ngu21a}. For instance, a two-stage data detection algorithm based on support vector machine was proposed in \cite{Ngu21a}, which performs close to ML when perfect CSI is available.

Although all the aforementioned works focus on designing low-complexity and accurate data detection methods to improve the system's performance, they use the original transmit constellation as a reference. The expected values of the soft-estimated symbols (after the linear receiver and prior to data detection) were characterized in \cite{Atz22,Atz21a} for a massive single-input multiple-output (SIMO) system with 1-bit ADCs. These expected values effectively capture the distortion in the transmit constellation due to 1-bit quantization during both the channel estimation and the uplink data transmission, which directly depends on the SNR. In this respect, it was demonstrated in \cite{Atz22} that the distortion at high SNR is such that the soft-estimated symbols resulting from the data symbols with the same phase become indistinguishable. Hence, the amplitude information cannot be recovered at very high SNR. Furthermore, simple single-UE data detection strategies based on the expected values of the soft-estimated symbols were proposed in \cite{Abd23} for independent and identically distributed (i.i.d.) Rayleigh fading.

\subsection{Contribution} \label{sec:I.B}

In contrast to the simplified system model in \cite{Atz22}, which characterized the expected values of the soft-estimated symbols for a single UE and i.i.d. Rayleigh fading, this paper investigates a more general and realistic multi-UE setting with correlated Rayleigh fading, which better captures the sparse channel characteristics between the UEs and the base station (BS) at high frequencies.
Additionally, while the analysis in \cite{Atz22} is limited to maximum ratio combining (MRC), we extend it to the more advanced ZF and minimum mean squared error (MMSE) receivers. In this context, we design a \textit{linear minimum mean dispersion} (LMMD) receiver tailored for data detection with 1-bit ADCs, which exploits the expected values of the soft-estimated symbols previously derived. Furthermore, we propose a joint data detection (JD) strategy that exploits the interdependence among the soft-estimated symbols of the interfering UEs, along with its low-complexity variant. 

The contributions of this paper are summarized as follows:
\begin{itemize}
    \item We characterize the expected values of the soft-estimated symbols for a multi-UE setting when the MRC receiver is adopted at the BS. The characterization of these expected values is of particular interest since they effectively capture the distortion in the transmit constellation due to 1-bit quantization during both the channel estimation and the uplink data transmission, which directly depends on the SNR. Our analysis shows that the expected value of the soft-estimated symbol of a specific UE depends on the SNR, the number of UEs, the number of BS antennas, the pilot matrix, and the data symbols transmitted by all the UEs. In particular, the interdependence among the soft-estimated symbols of the interfering UEs can be leveraged to improve the data detection performance.
    
    \item We extend the above analysis to scenarios where the ZF and MMSE receivers are adopted at the BS. Interestingly, the expected values of the soft-estimated symbols with ZF and MMSE can be asymptotically obtained by simply scaling their MRC counterparts, for which a closed-form expression was previously derived. We further derive this scaling factor in closed form.
    
    \item For the receiver design tailored for data detection with 1-bit ADCs, we account for the impact of the expected~values of the soft-estimated symbols. We propose the LMMD receiver that minimizes the mean dispersion of the soft-estimated symbols around the expected values obtained with the conventional receivers (MRC, ZF, and MMSE).
    
    \item For data detection with 1-bit ADCs, we propose new data detection strategies based on the minimum distance criterion with respect to the expected values of the soft-estimated symbols. In this respect, we begin by introducing UE-specific data detection strategies such as exhaustive, heuristic, and genie-aided data detection. These methods serve as intermediate steps to explore how the interdependence among the soft-estimated symbols of the interfering UEs can be leveraged to improve the SER. Then, we propose an enhanced multi-UE data detection strategy, referred to as JD, that considers parallel data detection over all the UEs and exploits the interdependence among their soft-estimated symbols. The computational complexity of JD is independent of the number of BS antennas and scales exponentially with the number of UEs, making it well suited for scenarios with small-to-moderate numbers of UEs served by a massive antenna array. Moreover, we present a low-complexity variant of JD obtained by reducing the size of the search space. Lastly, considering a simplified setting with i.i.d. Rayleigh fading, we derive an upper bound on the SER for JD and provide an exact SER expression for heuristic UE-specific data detection. 
 
    \item Numerical results show that the proposed LMMD receiver significantly outperforms the conventional receivers (MRC and MMSE) in terms of SER. Among the conventional receivers, MMSE achieves substantial performance gains over MRC due to the reduced dispersion of the soft-estimated symbols around their expected values. In addition, the proposed JD and its low-complexity variant greatly outperform UE-specific data detection by effectively capturing the interdependence among the soft-estimated symbols of the interfering UEs.
\end{itemize}

Part of this work was presented in our conference papers \cite{Rad23,Rad23a}. Specifically, \cite{Rad23} derived the closed-form expression of the expected values of the soft-estimated symbols obtained with MRC in a multi-UE setting and proposed UE-specific data detection strategies. On the other hand, \cite{Rad23a} provided a numerical evaluation of the expected values of the soft-estimated symbols obtained with the ZF and MMSE receivers, and introduced JD. In this paper, we provide a detailed proof of Theorem~\ref{thm:main}, which was previously presented in \cite{Rad23}, and analytically characterize the expected values of the soft-estimated symbols obtained with the ZF and MMSE receivers. Building on this, we propose a novel receiver design tailored for data detection with 1-bit ADCs. Moreover, we analyze the SER for the proposed JD and present comprehensive numerical results considering a variety of system parameters.

\smallskip

\textit{Outline.}
The rest of the paper is structured as follows. Section~\ref{sec:Sys} introduces the system model. Section~\ref{sec:Ex} characterizes the expected values of the soft-estimated symbols obtained with the conventional receivers. Section~\ref{sec:Beam} exploits this result to propose a new receiver design tailored for data detection with 1-bit ADCs. Section~\ref{sec:DD} presents novel data detection strategies and the SER analysis. Section~\ref{sec:NR} provides the numerical results and discussion. Lastly, Section~\ref{sec:Concl} concludes the paper.

\smallskip

\textit{Notation.}
$(\cdot)^\tran$, $(\cdot)^\herm$, and $(\cdot)^*$ represent transpose, Hermitian transpose, and conjugate, respectively. $\Re[\cdot]$ and $\Im[\cdot]$ denote the real and imaginary parts, respectively, whereas $j$ is the imaginary unit. $\mathbb{E}[\cdot]$ is the expectation operator. $\I_N$ and $\0_N$ denote the $N$-dimensional identity matrix and all-zero vector, respectively. $[\cdot]_{m,n}$ specifies the $(m,n)$th element of the matrix argument, whereas $x_{n}$ and $x_{\textrm{d},n}$ denote the $n$th elements of $\x$ and $\x_{\textrm{d}}$, respectively. $\{\cdot\}$ is used to denote sets. $\mathrm{vec}[\cdot]$ is the vectorization operator. $\otimes$ denotes the Kronecker product. $\sgn(\cdot)$ is the sign function. $\diag(\cdot)$ produces a diagonal matrix with the elements of the vector argument or the diagonal elements of the square matrix argument on its diagonal. $\blkdiag (\cdot)$ produces a block-diagonal matrix. $\mathcal{CN}(\0_N,\Sigmab)$ and $\mathcal{N}(\0_N,\Sigmab)$ are, respectively, the $N$-variate complex and real normal distributions with zero mean and covariance matrix $\Sigmab$. $\erf(x) \triangleq \frac{2}{\sqrt{\pi}} \int_{0}^{x} e^{-t^2} \diff t$ represents the error function for $x \in \Real$; for $x \in \Compl$, we have $\erf_{\textrm{c}}(x) \triangleq \erf \big(\Re[x] \big) +j \, \erf \big(\Im[x] \big)$.

\section{System Model} \label{sec:Sys}

Consider a single-cell massive MIMO system where a BS equipped with $M$ antennas serves $K$ single-antenna UEs in the uplink. Let $\H \triangleq [\h_{1}, \ldots, \h_{K}] \in \Compl^{M \times K}$ denote the uplink channel matrix, where $\h_{k}$ represents the channel vector of UE~$k$. Considering a general correlated Rayleigh fading channel model, we have $\h_{k} \sim \setC \setN (\0_{M}, \C_{\h_{k}}),~\forall k = 1, \ldots, K$, where $\C_{\h_{k}} \in \Compl^{M \times M}$ is the channel covariance matrix of UE~$k$. Furthermore, we define $\h \triangleq \mathrm{vec} [\H] \in \Compl^{M K}$ and, accordingly, we have $\h \sim \setC \setN (\0_{M K}, \C_{\h})$, with $\C_{\h} \triangleq \blkdiag (\C_{\h_{1}}, \ldots, \C_{\h_{K}}) \in \Compl^{M K \times M K}$. For simplicity and without loss of generality, we assume that all the UEs are subject to the same SNR $\rho$ during both the channel estimation and the uplink data transmission (as in, e.g., \cite{Li17,Jac17a,Atz22}). Each BS antenna is connected to two 1-bit ADCs, one for the in-phase and one for the quadrature component of the received signal. In this context, we introduce the 1-bit quantization function $Q(\cdot) : \Compl^{A \times B} \to \setQ$, with $\setQ \triangleq \sqrt{\frac{\rho K + 1}{2}} \{ \pm 1 \pm j \}^{A \times B}$~and
\begin{align} \label{eq:Q}
Q(\X) \triangleq \sqrt{\frac{\rho K + 1}{2}} \Big( \sgn \big( \Re[\X] \big) + j \, \sgn \big(\Im[\X] \big) \Big).
\end{align}

\subsection{Channel Estimation} \label{sec:SM_A}

As in \cite{Li17}, we utilize the Bussgang linear MMSE (BLMMSE) estimator to estimate the channels (we refer to \cite{Din25} for a discussion on its optimality). Let $\P \triangleq [\p_{1}, \ldots, \p_{K}] \in \Compl^{\tau \times K}$ denote the pilot matrix, where $\tau$ is the pilot length, $\p_{k} \in \Compl^{\tau}$ represents the pilot vector of UE~$k$, and $P_{u,k}$ is the $(u,k)$th element of $\P$. We assume $\tau \geq K$ and orthogonal pilots among the UEs. During the channel estimation, all the UEs simultaneously transmit their pilots, and the signal received at the input of the ADCs at the BS is given by
\begin{align} \label{eq:Yp}
\Y_{\mathrm{p}} \triangleq \sqrt{\rho} \H \P^{\herm} + \Z_{\mathrm{p}} \in \Compl^{M \times \tau}
\end{align}
where $\Z_{\mathrm{p}} \in \Compl^{M \times \tau}$ is a matrix of additive white Gaussian noise (AWGN) with i.i.d. $\setC \setN (0,1)$ elements. At this stage, we vectorize \eqref{eq:Yp} as
\begin{align}
\y_{\mathrm{p}} & \triangleq \mathrm{vec} [\Y_\mathrm{p}] = \sqrt{\rho} \bar{\P}^{*} \h + \z_\mathrm{p} \in \Compl^{M \tau}
\end{align}
with $\bar{\P} \triangleq \P \otimes \I_{M} \in \Compl^{M \tau \times M K}$ and $\z_\mathrm{p} \triangleq \mathrm{vec}[\Z_\mathrm{p}] \in \Compl^{M \tau}$. 
The BS observes the quantized signal
\begin{align} \label{rp}
\r_\mathrm{p} & \triangleq Q(\y_\mathrm{p}) \in \Compl^{M \tau}
\end{align}
and obtains an estimate of $\h$ via the BLMMSE estimator as~\cite{Li17}
\begin{align} \label{eq:h_hat}
\hat{\h} & \triangleq \sqrt{\rho} \C_\h \bar{\P}^\tran \A_\mathrm{p} \C_{\r_\mathrm{p}}^{-1} \r_\mathrm{p} \in \Compl^{M K}
\end{align}
with $\C_{\r_\mathrm{p}} \triangleq \Exp [\r_\mathrm{p} \r_\mathrm{p}^\herm] \in \Compl^{M \tau \times M \tau}$ and where
\begin{align} \label{eq:Ap}
\A_\mathrm{p} \triangleq \sqrt{\frac{2}{\pi}(\rho K + 1)} \diag(\C_{\y_\mathrm{p}})^{-\frac{1}{2}} \in \Compl^{M \tau \times M \tau}
\end{align}
is the Bussgang gain matrix, with $\C_{\y_\mathrm{p}} \triangleq \Exp [\y_\mathrm{p} \y_\mathrm{p}^\herm] = \rho \bar{\P}^* \C_\h\bar{\P}^\tran + \I_{M\tau} \in \Compl^{M \tau \times M \tau}$. Finally, the estimate of $\H$ is expressed as $\hat{\H} \triangleq [\hat{\h}_{1}, \ldots, \hat{\h}_{K}] \in \Compl^{M \times K}$, with
\begin{align} \label{eq:h_hat_k}
\hat{\h}_k \triangleq \sqrt{\rho} \C_{\h_k} \bar{\P}_{k}^\tran \A_\mathrm{p} \C_{\r_\mathrm{p}}^{-1} \r_\mathrm{p} \in \Compl^{M}
\end{align}
and $\bar{\P}_{k} \triangleq \p_{k} \otimes \I_{M} \in \Compl^{M \tau \times M}$.

\subsection{Uplink Data Transmission} \label{sec:SM_B}

\begin{figure*}[t!]
\centering
\begin{subfigure}[c]{0.49\textwidth}
\centering
\begin{tikzpicture}

\begin{axis}[
	width=6.5cm,
	height=6.5cm,
	xmin=-150, xmax=150,
	ymin=-150, ymax=150,
	axis equal,
    xlabel={I},
    ylabel={Q},
    xlabel near ticks,
    ylabel near ticks,
    xtick={-150,-100,-50,0,50,100,150},
    ytick={-150,-100,-50,0,50,100,150},
    legend style={at={(0.9,0.99)}, anchor=north west},
    legend style={font=\scriptsize, inner sep=1pt, fill opacity=0.75, draw opacity=1, text opacity=1},
    legend cell align=left,
	grid=both,
	x label style={font=\footnotesize},
	y label style={font=\footnotesize},
	ticklabel style={font=\footnotesize},
	clip marker paths=true,
 title={$K = 2$, $M = 128$, $\tau=31$, $\rho = 0$~dB},
	title style={font=\scriptsize, yshift=-1mm},
]

\addplot[only marks, black, mark=x]
table[x=Rs, y=Is, col sep=comma]
{Figures/Data/scatter0_soft_0dB.txt};
\addlegendentry{$\hat{x}_{1}^{\mathrm{(MRC)}}$}

\addplot[only marks, red, mark=o, thick]
table[x=Ex_r, y=Ex_i, col sep=comma]
{Figures/Data/scatter0_Ex_0dB.txt};
\addlegendentry{$\mathsf{e}_{1}^{\mathrm{(MRC)}}$}

\end{axis}

\end{tikzpicture}
\vspace{-2mm}
\caption{$\rho = 0$~dB}
\label{fig:scatter0_a}
\end{subfigure}
\hfill
\begin{subfigure}[c]{0.49\textwidth}
\centering
\begin{tikzpicture}

\begin{axis}[
scaled ticks=base 10:-3,
 tick scale binop=\times,
	width=6.5cm,
	height=6.5cm,
	xmin=-1500, xmax=1500,
	ymin=-1500, ymax=1500,
	axis equal,
    xlabel={I},
    ylabel={Q},
    xlabel near ticks,
    ylabel near ticks,
    xtick={-1500,-1000,-500,0,500,1000,1500},
    ytick={-1500,-1000,-500,0,500,1000,1500},
    legend style={at={(0.9,0.99)}, anchor=north west},
    legend style={font=\scriptsize, inner sep=1pt, fill opacity=0.75, draw opacity=1, text opacity=1},
    legend cell align=left,
	grid=both,
	x label style={font=\footnotesize},
	y label style={font=\footnotesize},
	ticklabel style={font=\footnotesize},
	clip marker paths=true,
 title={\hspace{12mm} $K = 2$, $M = 128$, $\tau=31$, $\rho = 20$~dB},
	title style={font=\scriptsize, yshift=-1mm},
]

\addplot[only marks, black, mark=x]
table[x=Rs, y=Is, col sep=comma]
{Figures/Data/scatter0_soft_20dB.txt};
\addlegendentry{$\hat{x}_{1}^{\mathrm{(MRC)}}$}

\addplot[only marks, red, mark=o, thick]
table[x=Ex_r, y=Ex_i, col sep=comma]
{Figures/Data/scatter0_Ex_20dB.txt};
\addlegendentry{$\mathsf{e}_{1}^{\mathrm{(MRC)}}$}

\end{axis}

\end{tikzpicture}
\vspace{-2mm}
\caption{$\rho = 20$~dB}
\label{fig:scatter0_b}
\end{subfigure}
\vspace{3mm}
\caption{Soft-estimated symbols (black markers) and their expected values (red markers) of UE~$1$ for $x_{2} = \frac{1}{\sqrt{10}}(-1 + j)$; the MRC receiver is adopted at the BS.} \label{fig:scatter0}
\end{figure*}
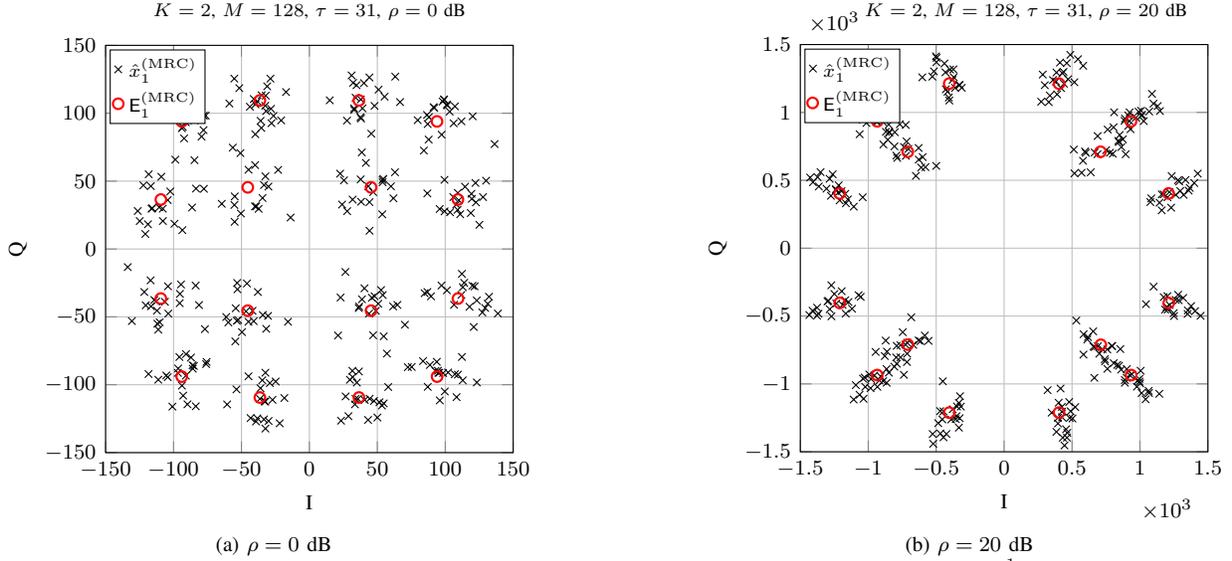

\begin{figure*}[t!]
\setcounter{equation}{18}
\begin{align}
\label{eq:Crp} [\C_{\r_\mathrm{p}}]_{(u-1)M+m,(v-1)M+n} & = \begin{cases}
\rho K + 1 & \quad \textrm{if}~m=n~\textrm{and}~u=v, \\
(\rho K + 1) \Big( \Omega \big( \Re [\zeta_{m,n,u,v}] \big) - j \, \Omega \big( \Im [\zeta_{m,n,u,v}] \big) \Big) & \quad \textrm{otherwise}
\end{cases}
\end{align}
\hrulefill
\vspace{-4mm}
\end{figure*}

Let $\x \triangleq [x_{1}, \ldots, x_{K}]^{\tran} \in \Compl^{K}$ denote the data symbol vector comprising the data symbols transmitted by the UEs. Moreover, let $\x \in \setS^{K}$, where $\setS \triangleq \{ s_{1}, \ldots, s_{L} \}$ represents the transmit constellation with $L$ data symbols. We use $\ell_{k} \in \{1, \ldots, L\}$ to denote the index of the data symbol from $\setS$ transmitted by UE~$k$. In this paper, we assume that $\setS$ corresponds to the 16-point quadrature amplitude modulation (QAM) constellation, i.e., $\setS = \frac{1}{\sqrt{10}} \big\{ \pm 1 \pm j, \pm 1 \pm j \, 3, \pm 3 \pm j, \pm 3 \pm j \, 3 \big\}$, which is normalized such that $\frac{1}{L} \sum_{l=1}^{L} |s_{l}|^{2} = 1$. Nonetheless, our analysis and data detection algorithms are general and can be applied to any transmit constellation. During the uplink data transmission, all the UEs simultaneously transmit their data symbols, and the signal received at the input of the ADCs at the BS is given by
\setcounter{equation}{7}
\begin{align} \label{eq:y}
\y \triangleq \sqrt{\rho} \H \x + \z \in \Compl^{M}
\end{align}
where $\z \sim \setC \setN (\0_{M}, \I_{M})$ is a vector of AWGN. The BS observes the quantized signal
\begin{align} \label{eq:r}
\r \triangleq Q(\y) \in \Compl^{M}
\end{align}
where we note that the scaling factor $\sqrt{\frac{\rho K + 1}{2}}$ in \eqref{eq:Q} is set such that the variance of $\r$ coincides with that of $\y$ (see, e.g., \cite{Jac17a,Atz22}). Then, the BS obtains a soft estimate of $\x$ via linear combining as
\begin{align}
\label{eq:x_hat} \hat{\x} & \triangleq [\hat{x}_{1}, \ldots, \hat{x}_{K}]^{\tran} = \V^{\herm} \r \in \Compl^{K}
\end{align}
where $\V \in \Compl^{M \times K}$ is the receiver matrix based on imperfect channel estimation, which is carried out as described in Section~\ref{sec:SM_A}. In this paper, we consider conventional receivers obtained by plugging the channel estimated with 1-bit ADCs into the well-known MRC, ZF, and MMSE structures, i.e.,
\begin{align}
\label{eq:V_MRC} \V^{\textrm{(MRC)}} & = \hat{\H}, \\
\label{eq:V_ZF} \V^{\textrm{(ZF)}} & = \hat{\H}(\hat{\H}^\herm\hat{\H})^{-1}, \\
\label{eq:V_MMSE} \V^{\textrm{(MMSE)}} & = (\rho \hat{\H}\hat{\H}^\herm + \I_{M})^{-1} \hat{\H}.
\end{align}
These conventional receivers, which are inherited from infinite-resolution systems, ignore the non-linear effects of the 1-bit ADCs. Therefore, in Section~\ref{sec:Beam}, we propose a novel receiver design tailored for data detection with 1-bit ADCs.

\section{Expected Values of the Soft-Estimated Symbols} \label{sec:Ex}

The prior work \cite{Atz22} focused on a single-UE setting with i.i.d. Rayleigh fading and analytically characterized the statistical properties (i.e., expected value and variance) of the soft-estimated symbols, which depend on the data symbol and pilot of the UE. In this paper, we consider a more general and realistic multi-UE setting with correlated Rayleigh fading. In this section, we derive a closed-form expression of the expected values of the soft-estimated symbols obtained with the MRC receiver and then connect it through asymptotic approximations to the cases of ZF and MMSE receivers. We show that these expected values depend on the data symbol vector $\x$ and pilot matrix $\P$.

\begin{figure*}[t!]
\centering
\begin{subfigure}[c]{0.59\textwidth}
\pgfdeclarelayer{background}
\pgfdeclarelayer{foreground}
\pgfsetlayers{background,main,foreground}

\begin{tikzpicture}

\begin{pgfonlayer}{background}
\begin{axis}[
	width=6.5cm,
	height=6.5cm,
	xmin=-130, xmax=130,
	ymin=-130, ymax=130,
	axis equal,
    xlabel={I},
    ylabel={Q},
    xlabel near ticks,
	ylabel near ticks,
    xtick={-120,-60,0,60,120},
    ytick={-120,-60,0,60,120},
    legend style={at={(0.01,0.99)}, anchor=north west},
	legend style={font=\scriptsize, inner sep=1pt, fill opacity=0.75, draw opacity=1, text opacity=1},
    legend cell align=left,
	grid=both,
	x label style={font=\footnotesize},
	y label style={font=\footnotesize},
	ticklabel style={font=\footnotesize},
	clip marker paths=true,
 title={$K = 2$, $M = 128$, $\tau=31$, $\rho = 0$~dB},
	title style={font=\scriptsize, yshift=-1mm},
]

\addplot[thick, blue, only marks, mark=o]
table[x=E_abs3_r, y=E_abs3_i, col sep=comma]
{Figures/Data/scatter1_abs3.txt};
\addlegendentry{$\mathsf{e}_{1}^{\mathrm{(MRC)}}$ for $x_2 \in \setS^{(1)}$}

\addplot[thick, black, only marks, mark=o]
table[x=E_abs2_r, y=E_abs2_i, col sep=comma]
{Figures/Data/scatter1_abs2.txt};
\addlegendentry{$\mathsf{e}_{1}^{\mathrm{(MRC)}}$ for $x_2 \in \setS^{(2)}$}

\addplot[thick, red, only marks, mark=o]
table[x=E_abs1_r, y=E_abs1_i, col sep=comma]
{Figures/Data/scatter1_abs1.txt};
\addlegendentry{$\mathsf{e}_{1}^{\mathrm{(MRC)}}$ for $x_2 \in \setS^{(3)}$}

\coordinate (zoom_coord1) at (120,-120);
\begin{scope}[>=latex]
\draw[->] (-84.5,28.3) -- (-1,-25);
\end{scope}

\draw[black] (axis cs:85,85) circle (0.5cm);

\begin{pgfonlayer}{foreground}
\begin{scope}[>=latex]
\draw[->] (axis cs:109,85) -- (axis cs:238.5,94);
\end{scope}
\end{pgfonlayer}

\coordinate (zoom_coord2) at (135,-5);
\coordinate (zoom_coord3) at (135,5);

\end{axis}
\end{pgfonlayer}

\begin{pgfonlayer}{foreground}
\begin{axis}[
	axis background/.style={fill=white},
	at={(zoom_coord1)},
	anchor={outer south east},
	width=3cm,
	height=3cm,
	xmin=-85.2, xmax=-84.6,
	ymin=28, ymax=28.6,
	axis equal,
	xtick={-85.2,-84.9,-84.6},
	ytick={28,28.3,28.6},
    grid=both,
	ticklabel style={font=\tiny},
	restrict x to domain=-85.2:-84.6,
	restrict y to domain=28:28.6,
	name=zoom_axis1,
]

\addplot[only marks, red, mark=o, thick] coordinates {(-84.6517443037143,28.29877180202)};

\addplot[only marks, red, mark=triangle, thick] coordinates {(-85.0954192971743,28.2608237135117)};

\addplot[only marks, red, mark=asterisk, thick] coordinates {(-85.030108269965,28.2444577181456)};

\addplot[only marks, red, mark=square, thick] coordinates {(-84.8058668721219,28.3936920464532)};

\end{axis}
\end{pgfonlayer}

\begin{pgfonlayer}{main}
\draw[black,fill=white] (zoom_axis1.outer south west) rectangle (zoom_axis1.outer north east);
\end{pgfonlayer}

\begin{pgfonlayer}{foreground}
\begin{axis}[
	axis background/.style={fill=white},
	at={(zoom_coord2)},
	anchor={outer north west},
	width=3.5cm,
	height=3.5cm,
	xmin=-1.5, xmax=1.5,
	ymin=-1.5, ymax=1.5,
	axis equal,
    xtick={-1.5,0,1.5},
    ytick={-1.5,0,1.5},
    grid=both,
	ticklabel style={font=\tiny},
	clip marker paths=true,
 title style = {align = center},
 title={Data symbols of UE~$2$},
	title style={font=\footnotesize},
	restrict x to domain=-1:1,
	restrict y to domain=-1:1,
	name=zoom_axis,
]
\addplot[only marks, red, mark=o, thick] coordinates {(-0.948683298050514,0.948683298050514)};
\addplot[only marks, red, mark=triangle, thick] coordinates {(-0.948683298050514,-0.948683298050514)};
\addplot[only marks, red, mark=asterisk, thick] coordinates {(0.948683298050514,0.948683298050514)};
\addplot[only marks, red, mark=square, thick] coordinates {(0.948683298050514,-0.948683298050514)};

\addplot[thick, black, only marks, mark=o]
table[x=xr, y=xi, col sep=comma]
{Figures/Data/QAM_abs2.txt};

\addplot[thick, blue, only marks, mark=o]
table[x=xr, y=xi, col sep=comma]
{Figures/Data/QAM_abs3.txt};

\end{axis}
\end{pgfonlayer}


\begin{pgfonlayer}{foreground}
\begin{axis}[
	axis background/.style={fill=none},
	at={(zoom_coord3)},
	anchor={outer south west},
	width=3.5cm,
	height=3.5cm,
	xmin=-1.5, xmax=1.5,
	ymin=-1.5, ymax=1.5,
	axis equal,
    xtick={-1.5,0,1.5},
    ytick={-1.5,0,1.5},
    grid=both,
	ticklabel style={font=\tiny},
	clip marker paths=true,
  title style = {align = center},
 title={Data symbols of UE~$1$},
	title style={font=\footnotesize},
	restrict x to domain=-1:1,
	restrict y to domain=-1:1,
	name=zoom_axis,
]

\addplot[thick, green, only marks, mark=o]
table[x=real_QAM, y=imag_QAM, col sep=comma]
{Figures/Data/QAM_points.txt};

\draw[black] (axis cs:0.95,0.95) circle (0.2cm);
\end{axis}
\end{pgfonlayer}


\end{tikzpicture} 
\vspace{-2mm}
\caption{$\rho = 0$~dB}
\label{fig:E1_a}
\end{subfigure}
\hfill
\begin{subfigure}[c]{0.39\textwidth}
\begin{tikzpicture}

\begin{axis}[
scaled ticks=base 10:-3,
 tick scale binop=\times,
	width=6.5cm,
	height=6.5cm,
	xmin=-1500, xmax=1500,
	ymin=-1500, ymax=1500,
	axis equal,
    xlabel={I},
    ylabel={Q},
    xlabel near ticks,
    ylabel near ticks,
    ylabel shift = -2pt,
    xtick={-1500,-1000,-500,0,500,1000,1500},
    ytick={-1500,-1000,-500,0,500,1000,1500},
    legend style={at={(0.01,0.99)}, anchor=north west},
	legend style={font=\scriptsize, inner sep=1pt, fill opacity=0.75, draw opacity=1, text opacity=1},
    legend cell align=left,
	grid=both,
	x label style={font=\footnotesize},
	y label style={font=\footnotesize},
	ticklabel style={font=\footnotesize},
	clip marker paths=true,
    title={\hspace{12mm} $K = 2$, $M = 128$, $\tau=31$, $\rho = 20$~dB},
	title style={font=\scriptsize, yshift=-1mm},
]

\addplot[thick, blue, only marks, mark=o]
table[x=E_abs3_r_20, y=E_abs3_i_20, col sep=comma]
{Figures/Data/scatter1_abs3.txt};
\addlegendentry{$\mathsf{e}_{1}^{\mathrm{(MRC)}}$ for $x_2 \in \setS^{(1)}$}

\addplot[thick, black, only marks, mark=o]
table[x=E_abs2_r_20, y=E_abs2_i_20, col sep=comma]
{Figures/Data/scatter1_abs2.txt};
\addlegendentry{$\mathsf{e}_{1}^{\mathrm{(MRC)}}$ for $x_2 \in \setS^{(2)}$}

\addplot[thick, red, only marks, mark=o]
table[x=E_abs1_r_20, y=E_abs1_i_20, col sep=comma]
{Figures/Data/scatter1_abs1.txt};
\addlegendentry{$\mathsf{e}_{1}^{\mathrm{(MRC)}}$ for $x_2 \in \setS^{(3)}$}

\end{axis}
\end{tikzpicture}
\vspace{-2mm}
\caption{$\rho = 20$~dB}
\label{fig:E1_b}
\end{subfigure}
\vspace{3mm}
\caption{Expected values of the soft-estimated symbols of UE~$1$ when UE~$2$ transmits all possible data symbols; the MRC receiver is adopted at the BS.}\label{fig:E1}
\end{figure*}

\subsection{Expected Values of the Soft-Estimated Symbols with MRC} \label{sec:Ex_A}

Here, we assume that the MRC receiver is adopted at the BS, where the receiver matrix is given in \eqref{eq:V_MRC}. Hence, the soft-estimated symbol of UE~$k$ can be expressed as $\hat{x}_{k}^{\textrm{(MRC)}} = \hat{\h}_{k}^{\herm} \r$, with $\hat{\h}_{k}$ and $\r$ given in \eqref{eq:h_hat_k} and \eqref{eq:r}, respectively. Let us define $\C_{\r \r_\mathrm{p}} \triangleq \Exp [\r \r_\mathrm{p}^{\herm}] \in \Compl^{M \times M \tau}$, which represents the cross-covariance matrix between the quantized signals received during the uplink data transmission and the channel estimation. Moreover, we introduce the function $\Omega(x) \triangleq \frac{2}{\pi} \arcsin(x)$ and the following preliminary definitions:
\begin{align}
\alpha_{m} & \triangleq \bigg[ \rho \sum_{k=1}^{K} \C_{\h_k} + \I_M \bigg]_{m,m}, \\
\beta_{m} & \triangleq \bigg[ \rho \sum_{k=1}^{K} \C_{\h_k} |x_{k}|^{2} + \I_M \bigg]_{m,m},
\end{align}
\begin{align}
\label{eq:zeta} \zeta_{m,n,u,v} & \triangleq \frac{\rho}{\sqrt{\alpha_{m} \alpha_{n}}} \bigg[ \sum_{k=1}^{K} \C_{\h_{k}}^{\tran} P_{u,k} P_{v,k}^{*} \bigg]_{m,n}, \\
\label{eq:eta} \eta_{m,n,u} & \triangleq \frac{\rho}{\sqrt{\alpha_{n} \beta_{m}}} \bigg[ \sum_{k=1}^{K} \C_{\h_{k}} x_{k} P_{u,k} \bigg]_{m,n}.
\end{align}
The following theorem provides a closed-form expression of the expected value of the soft-estimated symbol of UE~$k$ obtained with the MRC receiver for a given data symbol vector $\x$. This is denoted by $\mathsf{e}_{k}^{\textrm{(MRC)}}(\x,\P) \triangleq \Exp_{\H, \z, \z_\mathrm{p}} [\hat{x}_{k}^{\textrm{(MRC)}}]$. 

\begin{theorem} \label{thm:main}
Assuming that the MRC receiver is adopted at the BS, for a given data symbol vector $\x$ and pilot matrix $\P$, the expected value of the soft-estimated symbol of UE~$k$ is given by
\begin{align} \label{eq:Ex MRC}
\mathsf{e}_{k}^{\textnormal{(MRC)}}(\x,\P) = \sqrt{\rho} \tr( \C_{\r_\mathrm{p}}^{-1} \A_\mathrm{p} \bar{\P}_{k}^{*} \C_{\h_k} \C_{\r \r_\mathrm{p}})
\end{align}
where the ($(u-1)M+m,(v-1)M+n$)th element of $\C_{\r_\mathrm{p}}$ can be written as in \eqref{eq:Crp} at the top of the page and the ($m,(u-1)M+n$)th element of $\C_{\r \r_\mathrm{p}}$ can be written as
\setcounter{equation}{19}
\begin{align}
\nonumber [\C_{\r \r_\mathrm{p}}]_{m,(u-1)M+n} & = (\rho K + 1) \Big( \Omega \big( \Re [\eta_{m,n,u}] \big) \\
\label{eq:Crrp} & \phantom{=} \ + j \, \Omega \big( \Im [\eta_{m,n,u}] \big) \Big).
\end{align}
\end{theorem}

\begin{IEEEproof}
See Appendix~\ref{sec:App A}.
\end{IEEEproof}

\smallskip

\noindent The expression in \eqref{eq:Ex MRC} depends on: \textit{i)}~the specific data symbol vector $\x$ through $\C_{\r\r_\mathrm{p}}$ in \eqref{eq:Crrp} and $\eta_{m,n,u}$ in \eqref{eq:eta}; and \textit{ii)}~the pilot matrix $\P$ through $\C_{\r_\mathrm{p}}$ in \eqref{eq:Crp} and $\zeta_{m,n,u,v}$ in \eqref{eq:zeta} as well as $\C_{\r\r_\mathrm{p}}$ in \eqref{eq:Crrp} and $\eta_{m,n,u}$ in \eqref{eq:eta}. On the other hand, it does not depend on the specific channel realization but only on the channel covariance matrix $\C_{\h}$. A crucial practical aspect is that the expected value of the soft-estimated symbol of UE~$k$ also depends on the specific data symbols transmitted by the interfering UEs. This interdependence will be exploited to develop joint data detection strategies in Section~\ref{sec:DD}. To simplify the exposition in the rest of the paper, we omit the dependence on $\x$ and $\P$ and use the notation $\mathsf{e}_{k}^{\textrm{(MRC)}}$.

Considering $K = 2$ UEs, the soft-estimated symbols of UE~$1$ for each $x_{1} = s_{\ell_{1}} \in \setS$ are generated and plotted along with their expected values in Fig.~\ref{fig:scatter0} when the data symbol transmitted by the interfering UE is fixed to $x_{2} = \frac{1}{\sqrt{10}}(-1 + j)$. Two SNR values are considered, i.e., $\rho = 0$~dB (Fig.~\ref{fig:scatter0_a}) and $\rho = 20$~dB (Fig.~\ref{fig:scatter0_b}). Here, the soft-estimated symbols (black markers) originate from independent channel and AWGN realizations. We observe that, for each $x_{1} = s_{\ell_{1}} \in \setS$, the generated soft-estimated symbols are centered around the expected values obtained in \eqref{eq:Ex MRC} (red markers). In Fig.~\ref{fig:scatter0_b}, we further notice that the soft-estimated symbols and their expected values resulting from the data symbols with the same phase, i.e., $\frac{1}{\sqrt{10}} (\pm 1 \pm j)$ and $\frac{1}{\sqrt{10}} (\pm 3 \pm 3j)$, are almost overlapping due to the high SNR. This behavior, which was observed and analyzed in \cite{Atz22} for the single-UE case, implies that the amplitude information cannot be recovered at very high SNR. In general, the combination of 1-bit quantized channel estimation and uplink data transmission produces a clear distortion in the transmit constellation, which directly depends on the SNR.

Considering again $K = 2$ UEs, Fig.~\ref{fig:E1} depicts the expected values of the soft-estimated symbols of UE~$1$ when both UEs transmit all possible data symbols. As in Fig.~\ref{fig:scatter0}, two SNR values are considered, i.e., $\rho = 0$~dB (Fig.~\ref{fig:E1_a}) and $\rho = 20$~dB (Fig.~\ref{fig:E1_b}). The possible data symbols transmitted by UE~$1$ and UE~$2$ are also shown. For UE~$2$, the data symbols in the 16-QAM constellation with different amplitudes, i.e., $\setS^{(1)} \triangleq \frac{1}{\sqrt{10}} \{\pm 1 \pm j\}$, $\setS^{(2)} \triangleq \frac{1}{\sqrt{10}} \{\pm 1 \pm 3j,\pm 3 \pm 1j\}$, and $\setS^{(3)} \triangleq\frac{1}{\sqrt{10}} \{\pm 3 \pm 3j\}$, are indicated with different colors, i.e., blue, black, and red, respectively. Each red marker shown in the zoomed part in the bottom-right corner corresponds to a specific data symbol in $\setS^{(3)}$. In Fig.~\ref{fig:E1}, there are $16^2 = 256$ distinct pairs of data symbols transmitted by the two UEs, each corresponding to a different value of $\mathsf{e}_{1}^{\textrm{(MRC)}}$ in \eqref{eq:Ex MRC}. However, only $3 \times 16 = 48$ points can be clearly distinguished, which implies that there is significant overlap among many of the $256$ values of $\mathsf{e}_{1}^{\textrm{(MRC)}}$. This stems from the fact that, for a given $x_{1} = s_{\ell_{1}} \in \setS$, the data symbols with the same amplitude transmitted by UE~$2$ produce nearly the same value of $\mathsf{e}_{1}^{\textrm{(MRC)}}$. For this reason, the values of $\mathsf{e}_{1}^{\textrm{(MRC)}}$ are shown in blue, black, and red for $x_{2} \in \setS^{(1)}$, $x_{2} \in \setS^{(2)}$, and $x_{2} \in \setS^{(3)}$, respectively. To assess the impact of the SNR, we notice that, with $\rho = 0$~dB (Fig.~\ref{fig:E1_a}), all the groups of points are well separated. In contrast, with $\rho = 20$~dB (Fig.~\ref{fig:E1_b}), the groups of points corresponding to $x_{1}, x_{2} \in \setS^{(1)}$ roughly overlap with those corresponding to $x_{1}, x_{2} \in \setS^{(3)}$. This occurs because $x_{1}, x_{2} \in \setS^{(1)}$ have the smallest amplitude and are thus sensitive to the quantization distortion, which dominates at very high SNR.

\begin{figure}[t!]
\centering
\pgfdeclarelayer{background}
\pgfdeclarelayer{foreground}
\pgfsetlayers{background,main,foreground}

\begin{tikzpicture}

\begin{pgfonlayer}{background}
\begin{axis}[
	width=6.5cm,
	height=6.5cm,
	xmin=-130, xmax=130,
	ymin=-130, ymax=130,
	axis equal,
    xlabel={I},
    ylabel={Q},
    xlabel near ticks,
	ylabel near ticks,
    xtick={-120,-80,-40,0,40,80,120},
    ytick={-120,-80,-40,0,40,80,120},
    legend style={at={(0.01,0.99)}, anchor=north west},
    legend style={font=\scriptsize, inner sep=1pt, fill opacity=0.75, draw opacity=1, text opacity=1},
    legend cell align=left,
	grid=both,
	x label style={font=\footnotesize},
	y label style={font=\footnotesize},
	ticklabel style={font=\footnotesize},
	clip marker paths=true,
    title={$K = 3$, $M = 128$, $\tau=31$, $\rho = 0$~dB},
	title style={font=\scriptsize, yshift=-1mm},
]

\addplot[only marks, red, mark=o, thick]
table[x=Er, y=Ei, col sep=comma]
{Figures/Data/Ex_points_K3.txt};
\addlegendentry{$\mathsf{e}_{1}^{\mathrm{(MRC)}}$}

\addplot[only marks, green, mark=*, thick]
table[x=Er, y=Ei, col sep=comma]
{Figures/Data/Av_Ex_points_K3.txt};
\addlegendentry{$\bar{\mathsf{e}}_{1,\blue{\ell_{1}}}^{\mathrm{(MRC)}}$ (in \eqref{eq:E_kl})}

\coordinate (zoom_coord) at (axis cs:220,-125);
\draw[black] (axis cs:72,72) rectangle (axis cs:88,88) {};
\begin{scope}[>=latex]
\draw[->] (axis cs:82,74) -- (axis cs:120,17) {};
\end{scope}

\end{axis}
\end{pgfonlayer}

\begin{pgfonlayer}{foreground}
\begin{axis}[
	axis background/.style={fill=white},
	at={(zoom_coord)},
	anchor={outer south east},
	width=0.25\textwidth,
	height=0.25\textwidth,
	xmin=72, xmax=88,
	ymin=72, ymax=88,
	axis equal,
	ticks=none,
	restrict x to domain=72:88,
	restrict y to domain=72:88,
]

\addplot[only marks, red, mark=o, thick]
table[x=Er, y=Ei, col sep=comma]
{Figures/Data/Ex_points_K3.txt};

\addplot[only marks, green, mark=*, thick]
table[x=Er, y=Ei, col sep=comma]
{Figures/Data/Av_Ex_points_K3.txt};

\end{axis}
\end{pgfonlayer}

\end{tikzpicture}
\caption{Expected values of the soft-estimated symbols (red markers) and their average values for UE~$1$ (green markers) when UE~$2$ and UE~$3$ transmit all possible data symbols from $\setS$; the MRC receiver is adopted at the BS.} \label{fig:scatter_K=3}
\end{figure}
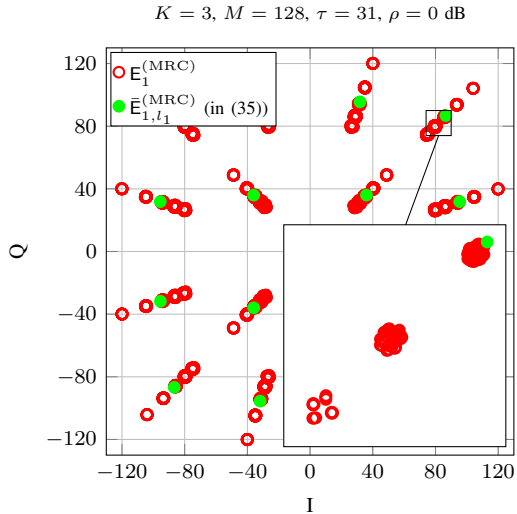

Fig.~\ref{fig:scatter_K=3} considers a similar setup but with $K = 3$ UEs. Here, there are $16^3 = 4096$ distinct triplets of data symbols transmitted by the three UEs, each corresponding to a different value of $\mathsf{e}_{1}^{\textrm{(MRC)}}$ in \eqref{eq:Ex MRC}. As in the case of $K = 2$ UEs shown in Fig.~\ref{fig:E1}, for a given $x_{1} = s_{\ell_{1}} \in \setS$, the data symbols with the same amplitude transmitted by UE~$2$ and UE~$3$ produce nearly the same value of $\mathsf{e}_{1}^{\textrm{(MRC)}}$. Interestingly, the dispersion of such values reduces as the pilot length increases since the channel estimates become more accurate. Note that the green markers in Fig.~\ref{fig:scatter_K=3} represent the average of the expected values of the soft-estimated symbols of UE~$1$ corresponding to a specific transmitted data symbol, as defined in \eqref{eq:E_kl}. These average values will be utilized in one of the data detection strategies presented in Section~\ref{sec:DD_A}.

\subsection{Expected Values of the Soft-Estimated Symbols with ZF and MMSE} \label{sec:EX_B}

We now assume that a conventional ZF or MMSE receiver is adopted at the BS. Let $\v_{k}^{\textrm{(ZF)}}$ and $\v_{k}^{\textrm{(MMSE)}}$ denote the $k$th columns of the ZF and MMSE receiver matrices given in \eqref{eq:V_ZF} and \eqref{eq:V_MMSE}, respectively. Then, the soft-estimated symbol of UE~$k$ can be expressed as $\hat{x}_{k}^{\textrm{(ZF)}} = (\v_{k}^{\textrm{(ZF)}})^\herm\r$ or $\hat{x}_{k}^{\textrm{(MMSE)}} = (\v_{k}^{\textrm{(MMSE)}})^\herm \r$, respectively, and the corresponding expected value of the soft-estimated symbol for a given data symbol vector $\x$ is given by
\begin{align}
\label{eq:Ex ZF} \mathsf{e}_{k}^{\textrm{(ZF)}} & \triangleq \mathbb{E}_{\H, \z, \z_\mathrm{p}}[\hat{x}_{k}^{\textrm{(ZF)}}], \\
\label{eq:Ex MMSE} \mathsf{e}_{k}^{\textrm{(MMSE)}} & \triangleq \mathbb{E}_{\H, \z, \z_\mathrm{p}}[\hat{x}_{k}^{\textrm{(MMSE)}}].
\end{align}
In the following lemma, we analyze the asymptotic orthogonality of the channel estimates of two independent UEs obtained with the BLMMSE estimator. This result will be used to derive closed-form approximations of \eqref{eq:Ex ZF} and \eqref{eq:Ex MMSE}.

\begin{lemma} \label{lemma1}
The channel estimates of two UEs $k$ and $k'$ obtained as in \eqref{eq:h_hat_k} are asymptotically orthogonal, i.e.,
\begin{align} \label{eq:as_orth}
\frac{\hat{\h}_k^\herm \hat{\h}_{k'}}{\sqrt{\mathbb E\big[\| \hat{\h}_k\|^2\big]} \sqrt{\mathbb E\big[\| \hat{\h}_{k'}\|^2\big]}} \to 0 \quad \textnormal{almost surely as} \ M \to \infty
\end{align}
if $\P$ is chosen such that $\P\P^\herm$ is circulant. However, \eqref{eq:as_orth} does not hold in general.
\end{lemma}

\begin{IEEEproof}
See Appendix~\ref{sec:App B}.
\end{IEEEproof}

\smallskip

According to Lemma~\ref{lemma1}, as $M \to \infty$, the normalized inner product between $\hat{\h}_k^\herm$ and $\hat{\h}_{k'}$ tends to its expected value, but this is generally not zero due to a bias in the channel estimation with 1-bit ADCs (even in the presence of orthogonal pilots). If $\P$ is chosen such that $\P\P^\herm$ is circulant, as in the case of discrete Fourier transform (DFT) pilot matrix, then the expected value is zero. In this case, the asymptotic orthogonality between the channel estimates is equivalent to the asymptotically favorable propagation property of massive MIMO channels \cite[Sec.~2.5.2]{massivemimobook}.

\begin{figure}[t!]
\centering
\begin{tikzpicture}

\begin{axis}[
	width=6.5cm,
	height=6.5cm,
	xmin=-1, xmax=1,
	ymin=-1, ymax=1,
	axis equal,
    xlabel={I},
    ylabel={Q},
    xlabel near ticks,
	ylabel near ticks,
        xtick={-0.8,-0.4,0,0.4,0.8},
    ytick={-0.8,-0.4,0,0.4,0.8},
    legend style={at={(0.01,0.99)}, anchor=north west},
    legend style={font=\scriptsize, inner sep=1pt, fill opacity=0.75, draw opacity=1, text opacity=1},
    legend cell align=left,
	grid=both,
	x label style={font=\footnotesize},
	y label style={font=\footnotesize},
	ticklabel style={font=\footnotesize},
	clip marker paths=true,
    title={$K = 2$, $M = 128$, $\tau=31$, $\rho = 0$~dB},
	title style={font=\scriptsize, yshift=-1mm},
]

\addplot[only marks, red, mark=o, thick]
table[x=Er_mmse_cl, y=Ei_mmse_cl, col sep=comma]
{Figures/Data/scatter_MMSE_closed.txt};
\addlegendentry{$\mathsf{e}_{1}^{\mathrm{(MMSE)}}$ (approx. in \eqref{eq:Ex MMSE app.})}

\addplot[only marks, blue, mark=star, thick]
table[x=Er_mmse, y=Ei_mmse, col sep=comma]
{Figures/Data/scatter_MMSE_closed.txt};
\addlegendentry{$\mathsf{e}_{1}^{\mathrm{(MMSE)}}$ (simulation)}

\end{axis}

\end{tikzpicture}
\caption{Expected values of the soft-estimated symbols of UE~$1$ when UE~$2$ transmits all possible data symbols; the MMSE receiver is adopted at the BS.} \label{Fig:scatter mmse closed}
\end{figure}
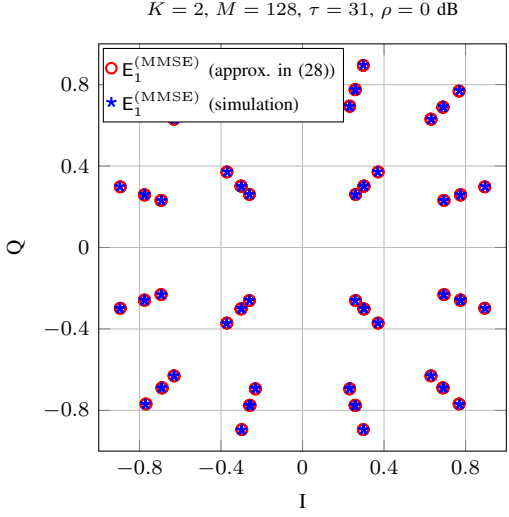

From Lemma~\ref{lemma1}, when $M$ is large, we have $\frac{\hat{\H}^\herm\hat{\H} }{M}\approx \diag \big( \big[ \frac{\|\hat{\h}_{1}\|^2}{M}, \ldots, \frac{\|\hat{\h}_{K}\|^2}{M} \big] \big)$ and thus we can write $\v_{k}^{\textrm{(ZF)}} \approx \frac{\hat{\h}_k}{\|\hat{\h}_{k}\|^2}$. Hence, \eqref{eq:Ex ZF} can be approximated as
\begin{align} \label{eq:Ex ZF app.}
     \mathsf{e}_{k}^{\textrm{(ZF)}} & \approx \mathbb{E}_{\H, \z, \z_\mathrm{p}}\bigg[\frac{\hat{\h}_k^\herm\r}{\|\hat{\h}_{k}\|^2}\bigg]\approx \frac{\mathsf{e}_{k}^{\textrm{(MRC)}}}{\mathbb{E}\big[\|\hat{\h}_{k}\|^2\big]}  
\end{align}
with
\begin{align}\label{eq:Eh_hat}
    \mathbb{E}\big[\|\hat{\h}_{k}\|^2\big] = \rho \tr(\C_{\h_k} \bar{\P}_k^\tran \A_{\mathrm{p}} \C_{\r_{\mathrm{p}}}^{-1} \A_{\mathrm{p}} \bar{\P}_k^* \C_{\h_k}).
\end{align}

To derive a closed-form approximation of \eqref{eq:Ex MMSE}, we first apply the matrix inversion lemma to write $(\rho \hat{\H}\hat{\H}^\herm + \I_{M})^{-1} = \B_{k}^{-1} - \rho \frac{\B_{k}^{-1} \hat{\h}_{k}\hat{\h}_{k}^\herm \B_{k}^{-1}}{1 + \rho \hat{\h}_{k}^\herm \B_{k}^{-1} \hat{\h}_{k}}$, with $\B_{k} \triangleq \rho\hat{\H}_{-k}\hat{\H}_{-k}^\herm  + \I_{M} \in \mathbb C^{M\times M}$ and $\hat{\H}_{-k} \triangleq [\hat{\h}_{1}, \ldots, \hat{\h}_{k-1}, \hat{\h}_{k+1}, \ldots, \hat{\h}_{K}] \in \mathbb C^{M\times (K-1)}$. For simplicity and without loss of generality, let us consider the case of $K = 2$ UEs and express the soft-estimated symbol of UE~$1$ as
\begin{align}\label{eq:x_1 MMSE}
    \hat{x}_{1}^{\textrm{(MMSE)}} = \hat{\h}_{1}^{\herm}\B_{1}^{-1}\r - \rho \hat{\h}_{1}^{\herm} \frac{\B_{1}^{-1} \hat{\h}_{1}\hat{\h}_{1}^\herm \B_{1}^{-1}}{1 + \rho \hat{\h}_{1}^\herm \B_{1}^{-1} \hat{\h}_{1}}\r.
\end{align}
From Lemma~\ref{lemma1}, when $M$ is large, we have
\begin{align}\label{eq:h1A}
    \hat{\h}_{1}^{\herm}\B_{1}^{-1} = \hat{\h}_{1}^{\herm} - \rho \hat{\h}_{1}^{\herm} \frac{\hat{\h}_{2}\hat{\h}_{2}^\herm}{1 + \rho {\|\hat{\h}_{2}\|}^2} \approx \hat{\h}_{1}^{\herm}.
\end{align}
Finally, \eqref{eq:Ex MMSE} can be approximated by plugging \eqref{eq:h1A} into \eqref{eq:x_1 MMSE}, which yields
\begin{align}\label{eq:Ex MMSE app.}
    \mathsf{e}_{k}^{\textrm{(MMSE)}} \approx \mathbb{E}_{\H, \z, \z_\mathrm{p}}\bigg[\frac{\hat{\h}_{k}^{\herm}\r}{1 + \rho \|\hat{\h}_{k}\|^2}\bigg] \approx  \frac{\mathsf{e}_{k}^{\textrm{(MRC)}}}{1 + \rho \mathbb{E}\big[\|\hat{\h}_{k}\|^2\big]}.
\end{align}

From \eqref{eq:Ex ZF app.} and \eqref{eq:Ex MMSE app.}, it is straightforward to observe that the expected values of the soft-estimated symbols obtained with the ZF and MMSE receivers asymptotically correspond to scaled versions of their MRC counterparts. Considering $K=2$ UEs, Fig.~\ref{Fig:scatter mmse closed} plots the expected values of the soft-estimated symbols of UE~$1$ when the MMSE receiver is adopted at the BS. We notice that the closed-form approximation in \eqref{eq:Ex MMSE app.} coincides with the Monte Carlo simulation even for a moderate number of BS antennas (i.e., $M = 128$). In general, this approximation is accurate when $\frac{M}{K}$ is sufficiently large.

\section{Receiver Design} \label{sec:Beam}

In the previous section, the receiver matrix was obtained by plugging the channel estimates (obtained with 1-bit ADCs) into the conventional MRC, ZF, and MMSE receiver structures. This type of receiver design, which is widely adopted in the literature on low-resolution massive MIMO \cite{Jac17a}, is inherited from infinite-resolution systems and thus ignores the non-linear effects of the low-resolution ADCs. Recently, a Bussgang-based MMSE receiver for 1-bit ADCs was proposed in \cite{Ngu21}. However, this receiver is designed under the assumption that the received signal in \eqref{eq:y} is approximately Gaussian, which does not hold for a small-to-moderate number of UEs. Consequently, it overlooks the distorted shape of the transmit constellation due to the 1-bit ADCs, which is instead captured by the expected values of the soft-estimated symbols derived in Section~\ref{sec:Ex}. Therefore, we propose a new receiver design that exploits these expected values of the soft-estimated symbols.

We begin by formulating the receiver design problem as
\begin{align}
\label{eq:V design (1)}
    \V^{\star} &=
\argmin_{\V} \mathbb{E}_{\x,\z}\Big[\big\|\V^\herm\r - \mathbb {E}_{\H,\z,\z_\mathrm{p}}[\V^\herm\r|\x]\big\|^2 \Big]
\end{align}
where $\mathbb E_{\x}[\cdot]$ is a discrete expectation over the $L^K$ possible data symbol vectors. The problem in \eqref{eq:V design (1)} aims at deriving a receiver, termed as \textit{linear minimum mean dispersion} (LMMD) receiver, that minimizes the mean dispersion of the soft-estimated symbols (outer expectation) around their expected values (inner expectation). To make the formulation more tractable, we impose a conventional receiver structure in the inner expectation that fixes the points around which the mean dispersion of the soft-estimated symbols is minimized. In this regard, we showed in Section~\ref{sec:EX_B} that the expected values of the soft-estimated symbols for the MRC, ZF, and MMSE receivers are asymptotically identical, differing only by a scaling factor. Therefore, we impose the MRC receiver in the inner expectation, so that \eqref{eq:V design (1)} can be reformulated as\footnote{Imposing a specific receiver structure in the inner expectation of \eqref{eq:V design (1)} results in a suboptimal strategy and, hence, \eqref{eq:V design (2)} is not equivalent to \eqref{eq:V design (1)}. However, it provides a tractable formulation and leads to significantly improved data detection performance compared with the conventional receivers.}
\begin{align}
\label{eq:V design (2)}
    \hspace{-2mm} \V^{(\mathrm{LMMD})} &= \argmin_{\V} \mathbb{E}_{\x,\z}\Big[\big\|\V^\herm\r - \mathbb {E}_{\H,\z,\z_\mathrm{p}} [\hat{\H}^\herm\r|\x] \big\|^2 \Big].
\end{align}
Finally, after some manipulations, the LMMD receiver can be obtained by setting the derivative of the objective of \eqref{eq:V design (2)} to zero, which yields
\begin{align}
\label{eq:WLMMD}
    \V^{(\mathrm{LMMD})} &= \frac{\sqrt{\rho}}{L^K} \C_{\r}^{\dagger}\sum_{\x \in \setS^K} \G(\x) \hat{\H} \x{\e(\x)}^\herm 
\end{align}
with
\begin{align}
    \e(\x) & \triangleq \mathbb{E}_{\H,\z,\z_\mathrm{p}} [\hat{\H}^\herm\r|\x] \\
    & = \big[ \mathsf{e}_{1}^{\textrm{(MRC)}} (\x, \P), \ldots, \mathsf{e}_{K}^{\textrm{(MRC)}} (\x, \P) \big]^\tran \! \in \Compl^K \label{eq:e(x)}
\end{align}
where $\mathsf{e}_{k}^{\textrm{(MRC)}} (\x, \P)$ is given in \eqref{eq:Ex MRC}. Furthermore, we have $\C_{\r} \triangleq \Exp_{\x,\z} [\r \r^{\herm}]\in \mathbb{C}^{M\times M}$ and
\begin{align}
\label{eq:G(x)}
  \G(\x) &= \C_{\y\r|\x}^\herm \C_{\y|\x}^{-1} \in \mathbb{C}^{M\times M}
\end{align}
is the Bussgang gain matrix, with $\C_{\y|\x} \triangleq \mathbb E_{\z}[\y\y^\herm|\x] \in \mathbb{C}^{M\times M}$ and $\C_{\y\r|\x} \triangleq \mathbb{E}_{\z}[\y \r^\herm|\x] 
\in \mathbb{C}^{M\times M}$. The detailed derivations are provided in Appendix~\ref{sec:App C}.

Some observations are in order. First, $\V^{(\mathrm{LMMD})}$ in \eqref{eq:WLMMD} needs to be computed for each channel realization, as the conventional receivers in \eqref{eq:V_MRC}--\eqref{eq:V_MMSE}. Specifically, the auto-correlation matrix $\C_{\y|\x} = \rho \H\x\x^\herm\H^\herm + \I_M$ in \eqref{eq:C_{y|x}} depends on the specific channel realization, and the same holds for $\C_{\r}$ (see \eqref{eq:Cr|x (2) (App)}--\eqref{eq:E[rr^H] (App)}) and $\C_{\y\r|\x}$ (see \eqref{eq:Cyr|x (2) (App)}), which are detailed in Appendix~\ref{sec:App C}. The practical computation of these the auto-correlation matrices under imperfect CSI follows from replacing the channel matrix $\H$ with its estimate $\hat{\H}$.
 
Hence, all the auto-correlation matrices used in \eqref{eq:WLMMD} can be readily estimated based on the channel estimate. Second, we remark that the expected values of the soft-estimated symbols in $\e(\x)$ need not be computed for each channel realization as they depend only on the channel covariance matrix $\C_{\h}$. The computational complexity of the LMMD receiver lies in the pre-processing stage for each channel realization and grows exponentially with $K$ due to the summation in \eqref{eq:WLMMD}. The LMMD receiver is thus suitable for small-to-moderate numbers of UEs, as considered in this~paper.

Note that the channel estimation error with 1-bit ADCs increases at high SNR \cite{Atz22}. Hence, imperfect CSI can significantly deteriorate the performance of the LMMD receiver when the SNR is set too high. Nonetheless, it is shown in Section~\ref{sec:NR} that the LMMD receiver significantly outperforms the conventional receivers at the optimal SNR, which can be achieved via UE power control \cite{Rav23}.

\section{Data Detection Strategies}\label{sec:DD}

In this section, we propose efficient data detection strategies that leverage the interdependent expected values of the soft-estimated symbols, described in Section~\ref{sec:Ex}, alongside the minimum distance criterion to map each soft-estimated symbol in \eqref{eq:x_hat} to one of the possible data symbols in $\setS$. In this paper, the soft-estimated symbols are obtained with the conventional receivers (MRC, ZF, and MMSE) or with the LMMD receiver proposed in Section~\ref{sec:Beam}, which also builds on the results of Section~\ref{sec:Ex}. To reduce the computational complexity, the two problems of obtaining and mapping the soft-estimated symbols are addressed separately in this paper. As a result, the computational complexity of the proposed data detection strategies mainly depends on the size of the transmit constellation and the number of UEs. In contrast, end-to-end solutions that perform data detection directly from the 1-bit quantized received signal, such as ML, near ML \cite{Cho16}, and RML \cite{Ngu21}, exhibit computational complexity that scales with the number of BS antennas and do not necessarily lead to better performance, as demonstrated in Section~\ref{sec:NR}.

The data detection strategies proposed next can be applied to any receiver structure described in Sections~\ref{sec:Ex} and~\ref{sec:Beam}. In this regard, we use $\mathsf{e}_{k}$ to denote the expected value of the soft-estimated symbol of UE~$k$ obtained by any of the conventional receivers. In the following, we use $\x_{-k}\triangleq[x_{1}, \ldots, x_{k-1}, x_{k+1}, \ldots, x_{K}]^\tran \in \Compl^{K-1}$ to denote the data symbol vector comprising the data symbols transmitted by the interfering UEs. Furthermore, $l_{k} \in \{1, \ldots, L\}$ and $l_{k}^{\star}$ denote, respectively, the search index and the detected index of the data symbol of UE~$k$. A detection error occurs if $l_{k}^{\star} \neq \ell_{k}$.

\subsection{UE-Specific Data Detection Strategies}\label{sec:DD_A}

Here, we consider the soft-estimated symbols and their expected values of each target UE separately. In this respect, we present three UE-specific data detection strategies: \textit{1)} exhaustive, \textit{2)} heuristic, and \textit{3)} genie-aided data detection.

\smallskip

\textbf{\textit{Exhaustive UE-specific data detection (E-USD).}} This strategy uses the statistical information (e.g., covariance matrices) of the interfering UEs to detect the data symbol of the target UE. Let $\setE_{k} \triangleq \{ \mathsf{e}_{k}, \forall \x \in \setS^{K} \}$ denote the set of the expected values of the soft-estimated symbols for UE~$k$ obtained from all possible data symbol vectors, with $|\setE_{k}| = L^{K}$. In E-USD, the soft-estimated symbol of UE~$k$ is mapped to one of the expected values in $\setE_{k}$ as
\begin{align}
\mathsf{e}_{k}^{\star} = \argmin_{\mathsf{e}_{k} \in \setE_{k}} |\hat{x}_{k} - \mathsf{e}_{k}|
\end{align}
from which $l_{k}^{\star}$ can be readily extracted. This strategy amounts to performing an exhaustive search over all the $L^{K}$ possible values of $\mathsf{e}_{k}$, with complexity increasing exponentially with~$K$.

\smallskip

\textbf{\textit{Heuristic UE-specific data detection (H-USD).}} This strategy considers the average of the expected values of the soft-estimated symbols of the target UE over all possible data symbols transmitted by the interfering UEs. Let $\setE_{k,l_{k}} \triangleq \{ \mathsf{e}_{k} : x_{k} = s_{l_{k}},~\forall \x_{-k} \in \setS^{K-1} \} \subset \setE_{k}$ be the set containing the elements in $\setE_{k}$ corresponding to $x_{k} = s_{l_{k}}$, with $|\setE_{k,l_{k}}| = L^{K-1}$. Furthermore, let us define
\begin{align} \label{eq:E_kl}
\bar{\mathsf{e}}_{k,l_{k}} \triangleq \frac{1}{L^{K-1}} \sum_{t \in \setE_{k,l_{k}}} t
\end{align}
which represents the average of expected values of the soft-estimated symbols of the target UE over all possible data symbols transmitted by the interfering UEs (corresponding to the green markers in Fig.~\ref{fig:scatter_K=3}). In H-USD, the index of the detected data symbol of UE~$k$ is obtained as
\begin{align}\label{eq:l_k^star H-USD}
l_{k}^{\star} = \argmin_{l_{k} \in \{1, \ldots, L\}} |\hat{x}_{k} - \bar{\mathsf{e}}_{k,l_{k}}|.
\end{align}
This strategy is a heuristic, low-complexity variant of E-USD, reducing the size of the search space from $L^K$ to $L$.

\smallskip

\textbf{\textit{Genie-aided data detection.}} This strategy assumes that a genie ideally provides the data symbols transmitted by the interfering UEs to minimize the uncertainty in detecting the data symbol of the target UE. Hence, for UE~$k$, $\x_{-k}$ is assumed to be perfectly known, which reduces the size of the search space from $L^K$ to $L$. While clearly impractical, this strategy is considered solely for the purpose of evaluating the impact of perfectly known data symbols transmitted by the interfering UEs on the data detection performance of the target UE.

\subsection{Enhanced Multi-UE Data Detection Strategies} \label{sec:DD_B}

E-USD proposed in Section~\ref{sec:DD_A} maps each soft-estimated symbol of the target UE to one of the $L^{K}$ expected values of the soft-estimated symbols based on the minimum distance criterion. This method is highly inefficient since it treats each UE individually without leveraging the interdependence among the soft-estimated symbols of the interfering UEs. To address these issues, we propose two data detection strategies that exploit this interdependence: \textit{1)} joint data detection (JD) and \textit{2)} $N$-point joint data detection ($N$-JD). The proposed JD and $N$-JD strategies use the minimum distance criterion to map each soft-estimated symbol in \eqref{eq:x_hat} to one of the expected values of the soft-estimated symbols, which are detailed in Section~\ref{sec:Ex}, and thus to one of the possible data symbols in $\setS$.

\smallskip

\textbf{\textit{Joint data detection (JD).}} This strategy takes the expected values of the soft-estimated symbols of the interfering UEs into account, which leads to enhanced multi-UE data detection of the target UE. Regardless of which receiver is adopted at the BS, let $\mathsf{e}_{k,l_{k}} (\x_{-k})$ denote the expected value of the soft-estimated symbol of UE~$k$ for $x_{k} = s_{l_{k}}$ when the other UEs transmit $\x_{-k}$. For simplicity, and without loss of generality, we consider $K = 2$ in the following. Let $\boldsymbol{\mathsf{e}}_{l_{1},l_{2}}\triangleq [\mathsf{e}_{1,l_{1}} (s_{l_{2}}),\mathsf{e}_{2,l_{2}} (s_{l_{1}})]^\tran$  denote the vector containing the expected values of the soft-estimated symbols of the two UEs for $x_{1} = s_{l_{1}}$ and $x_{2} = s_{l_{2}}$. Let $\setE \triangleq \{ \boldsymbol{\mathsf{e}}_{l_{1},l_{2}}, \forall (s_{l_{1}},s_{l_{2}}) \in \setS^{2} \}$ denote the set of vectors comprising the expected values of the soft-estimated symbols for both UEs resulting from all possible data symbol vectors, with $|\setE| = L^{2}$. In JD, the soft-estimated symbol vector $\hat{\x}$ is mapped to one of the vectors in $\setE$ as
\begin{align}
\boldsymbol{\mathsf{e}}^{\star} = \argmin_{\boldsymbol{\mathsf{e}}_{l_{1},l_{2}} \in \setE} \|\hat{\x} - \boldsymbol{\mathsf{e}}_{l_{1},l_{2}}\|
\end{align}
from which $\{l_{k}^{\star}\}_{k=1}^{K}$ can be extracted. In the general case, this strategy amounts to performing an exhaustive search over all the $L^{K}$ possible vectors $\boldsymbol{\mathsf{e}}_{l_{1},\ldots,l_{K}}\triangleq \big[ \mathsf{e}_{1,l_{1}} (\x_{-1}),\ldots,\mathsf{e}_{K,l_{K}} (\x_{-K}) \big]^\tran$. Hence, the complexity of this strategy increases exponentially with $K$ as for E-USD.

\smallskip

\begin{figure*}[t!]
\setcounter{equation}{42}
\begin{align}\label{eq:Cz' JD}
    \nonumber \mathbb{E}\big[\Re\big[\hat{\x}^\herm \f_t(\x)\big]\Re\big[\f_{t'}(\x)^\herm\hat{\x}\big]\big] &=   \tr\big(\Re\big[\f_{t}(\x)\big]\Re\big[\f_{t'}(\x)\big]^\tran \C_{\hat{\x}}^{(\Re,\Re)}(\x)\big) + \tr\big(\Re\big[\f_{t}(\x)\big]\Im\big[\f_{t'}(\x)\big]^\tran \C_{\hat{\x}}^{(\Im,\Re)}(\x)\big) \\& 
   \phantom{=} \ + \tr\big(\Im\big[\f_{t}(\x)\big]\Re\big[\f_{t'}(\x)\big]^\tran \C_{\hat{\x}}^{(\Re,\Im)}(\x)\big) + \tr\big(\Im\big[\f_{t}(\x)\big]\Im\big[\f_{t'}(\x)\big]^\tran \C_{\hat{\x}}^{(\Im,\Im)}(\x)\big)
\end{align}

\vspace{-3mm}

\hrulefill
\begin{align}       
\overline{\mathsf{P}}_{k}^{\textrm{(JD)}}(\x) &= 1 - \frac{1}{(2\pi)^{\frac{(L-1)L^{K-1}}{2}} |\boldsymbol{\Sigma}^{(\textrm{JD})}(\x)|^\frac{1}{2}}\int_{\mathbb{R}_{+}^{(L-1)L^{K-1}}}\exp^{-\frac{1}{2}(\u-\mub^{(\textrm{JD})}(\x))^\tran \boldsymbol{\Sigma}^{(\textrm{JD})}(\x)^{-1}(\u-\mub^{(\textrm{JD})}(\x))} \diff \u \label{eq:Pr JD}
\end{align}
\hrulefill
\vspace{-4mm}
\end{figure*}

\textbf{\textit{$N$-point joint data detection ($N$-JD).}} This strategy can be seen as a low-complexity variant of JD that considers the $N \leq L$ values of $\bar{\mathsf{e}}_{k,l_{k}}$ in \eqref{eq:E_kl} that are closest to each UE's soft-estimated symbol $\hat{x}_{k}$. From \eqref{eq:l_k^star H-USD}, let $l_{k}^{(i)}$ denote the index of the detected data symbol of UE~$k$, which corresponds to the $i$th smallest $|\hat{x}_{k} - \bar{\mathsf{e}}_{k,l_{k}^{(i)}}|$, with $i \in \{1, \ldots, N\}$, and let $\setS'_{k} = \{s_{l_{k}^{(i)}}\}_{i=1}^{N}$. Recalling the definition of $\boldsymbol{\mathsf{e}}_{l_{1},\ldots,l_{K}}$ introduced above for JD, let $\setE' \triangleq \{ \boldsymbol{\mathsf{e}}_{l_{1},\ldots,l_{K}}, \forall (s_{l_{1}},\ldots,s_{l_{K}}) \in \prod_{k=1}^{K} \setS'_{k} \}$ denote the restricted set of vectors $\boldsymbol{\mathsf{e}}_{l_{1},\ldots,l_{K}}$ resulting from the data symbol vectors belonging to the Cartesian product of $\setS'_{k}$ across all the UEs, with $|\setE'| = N^{K}$. In $N$-JD, the soft-estimated symbol vector $\hat{\x}$ is mapped to one of the vectors in $\setE'$ as
\setcounter{equation}{38}
\begin{align}
\boldsymbol{\mathsf{e}}^{\star} = \argmin_{\boldsymbol{\mathsf{e}}_{l_{1},\ldots,l_{K}} \in \setE'}\|\hat{\x} - \boldsymbol{\mathsf{e}}_{l_{1},\ldots,l_{K}}\|
\end{align}
from which $\{l_{k}^{\star}\}_{k=1}^{K}$ can be readily extracted. As a result, the size of the search space is $N^K$, which can be made considerably smaller than $L^{K}$ by appropriately selecting the value of $N$. Lastly, we note that $N$-JD with $N=1$ corresponds to H-USD presented in Section~\ref{sec:DD_A}.

The computational complexity of the proposed enhanced multi-UE data detection strategies is independent of $M$ but grows exponentially with $K$, making them well suited for scenarios with small-to-moderate numbers of UEs served by a massive antenna array. For increasing $K$, the gains from joint data detection diminish, while the computational complexity becomes prohibitive. This is because the interference from several UEs tends to be approximately Gaussian (as in the assumption of \cite{Ngu21}) and the expected values of the soft-estimated symbols increasingly overlap (cf. Fig.~\ref{fig:E1} and Fig.~\ref{fig:scatter_K=3}). Consequently, for large $K$, leveraging the interdependence among the soft-estimated symbols of the interfering UEs offers minimal benefit. Nevertheless, scenarios with small-to-moderate numbers of UEs naturally arise in practice, e.g., in mmWave and sub-THz systems, where the number of UEs scheduled per resource block is typically limited due to the use of highly directional beams and the reduced coverage area \cite{Atz25}.

\renewcommand{\arraystretch}{1.25}
\begin{table}[t!]
\footnotesize
\centering
\begin{tabular}{|c|c|c|}
\hline
\multirow{2}{*}{\makecell{\textbf{Data detection} \\ \textbf{strategy}}} & \multicolumn{2}{c|}{\textbf{Computational complexity}} \\ 
\cline{2-3}
                                                & $\mathcal{O}(\cdot)$ notation     & FLOPs \\
\hline
\makecell{RML with \\ NN search \cite{Ngu21}}   & $\mathcal{O}(N_{\mathrm{NN}}KM)$  & \makecell{$2N_{\mathrm{NN}}M (\epsilon + 4K-1)$ \\$+N_{\mathrm{NN}}-1$} \\
\hline
E-USD                                           & $\mathcal{O}(L^K)$                & $6L^K -1$ \\
\hline
H-USD                                           & $\mathcal{O}(L)$                  & $6L-1$ \\
\hline
\makecell{Genie-aided \\ data detection}        & $\mathcal{O}(L)$                  & $6L-1$ \\
\hline
JD                                              & $\mathcal{O}(L^K)$                & $6KL^K-1$ \\
\hline
$N$-JD                                          & $\mathcal{O}(N^K)$                & \makecell{$\big(N\mathrm{log}(L) + 7L\big)K$ \\$ + 6KN^K -1$} \\
\hline
\end{tabular}
\caption{Computational complexity in $\mathcal{O}(\cdot)$ notation and FLOPs.} \label{Table:I}
\end{table}

\subsection{Computational Complexity Analysis} \label{sec:DD_C}

The computational complexity of all the proposed data detection strategies is summarized in Table~\ref{Table:I}.
The RML method with NN search in \cite{Ngu21}, which is considered as a baseline in Section~\ref{sec:NR}, is also included. This technique identifies a set of $N_{\mathrm{NN}}$ data symbol vectors in $\setS ^K$ that are closest to $\hat{\x}$ and then performs a search within this reduced set. In this regard, while the computational complexity of RML does not grow exponentially with the number of UEs, it depends on the number of BS antennas and can become significant when $M$ is large. On the other hand, the computational complexity of the proposed JD is $\mathcal{O}(L^K)$. Lastly, the computational complexity of the proposed $N$-JD is $\mathcal{O}(N^K)$, which is the lowest among the enhanced multi-UE data detection strategies. As discussed in Section~\ref{sec:DD_B}, while this complexity still grows exponentially with $K$, $N$ is adjustable and even very small values of $N$ (as low as 3 or 4) can deliver remarkable performance. Thus, the complexity of $N$-JD is significantly lower than that of both JD and RML, particularly for small-to-moderate values of $K$ and large values of $M$. In Table~\ref{Table:I}, we further provide the computational complexity in terms of floating point operations (FLOPs). In this context, we consider heap-based partial sorting for the sorting operations of $N$-JD. Moreover, $\epsilon$ represents the total number of FLOPs required for the logarithm and exponential functions, which depends on the implementation algorithm and the desired precision.

\subsection{SER Analysis} \label{sec:DD_D}

Providing a closed-form analysis of the SER in the presence of 1-bit ADCs is generally a formidable task, as it involves the orthant probability of a quadratic function of a multivariate normal random variable. Focusing on 1-bit SIMO systems with i.i.d. Rayleigh fading, \cite{Abd23} derived an approximated upper bound on the SER using numerous simplifications, whereas \cite{Rav25} provided a rigorous SER analysis for quadrature phase-shift keying (PSK) modulation. By considering a simplified setting with i.i.d. Rayleigh fading, we are able to derive an upper bound on the SER for JD and the exact SER for H-USD, both based on integrals that can be evaluated numerically with moderate computational effort. For notational simplicity, in this section, we replace the group index $l_1,\ldots,l_K$ with a single index $t$ and denote $\boldsymbol{\mathsf{e}}_{l_1,\ldots,l_K}$ as $\boldsymbol{\mathsf{e}}_{t}$. Let $\mathcal{T}$ be the set of all possible combinations of data symbol indices of the $K$~UEs with $l_k \neq \ell_k$, with $|\mathcal{T}| = (L-1)L^{K-1}$: this comprises all group indices leading to a detection error for UE~$k$. In this setting, $\boldsymbol{\mathsf{e}}_{t}$ with $t \in \mathcal{T}$ represents a vector of the expected values of the soft-estimated symbols with incorrect data symbol of UE~$k$; in contrast, $\boldsymbol{\mathsf{e}}_{\ell_1,\ldots,\ell_K}$ is the vector of the expected values corresponding to the data symbols transmitted by the $K$~UEs.

Considering JD, let $\xib^{(\textrm{JD})} \in \mathbb{R}^{(L-1)L^{K-1}}$ denote the detection error vector. The $t$th element of $\xib^{(\textrm{JD})}$, with $t \in \mathcal{T}$, is defined~as
\begin{align}\label{eq:xi JD (1)}
    [\xib^{(\textrm{JD})}]_t &\triangleq \|\hat{\x} - \boldsymbol{{\mathsf{e}}}_{t} \|^2 -  \|\hat{\x} - \boldsymbol{{\mathsf{e}}}_{\ell_{1},\ldots,\ell_{K}} \|^2
\end{align}
where a negative element in $\xib^{(\textrm{JD})}$ indicates a detection error. As shown in Appendix~\ref{sec:App E_A}, $\xib^{(\textrm{JD})}$ approximately follows $\mathcal{N}\big(\mub^{(\textrm{JD})}(\x),\boldsymbol{\Sigma}^{(\textrm{JD})}(\x)\big)$ when $M$ is large, with $\mub^{(\textrm{JD})}(\x) \triangleq \mathbb{E}[\xib^{(\textrm{JD})}|\x] \in \mathbb{R}^{(L-1)L^{K-1}}$ and $\boldsymbol{\Sigma}^{(\textrm{JD})}(\x) \triangleq \mathbb{E}\big[\xib^{(\textrm{JD})}(\xib^{(\textrm{JD})})^\tran|\x \big] - \mub^{(\textrm{JD})}(\x)(\mub^{(\textrm{JD})}(\x))^\tran  \in \mathbb{R}^{((L-1)L^{K-1})\times ((L-1)L^{K-1})}$. The $t$th element of $\mub^{(\textrm{JD})}(\x)$ and the $(t,t')$th element of $\boldsymbol{\Sigma}^{(\textrm{JD})}(\x)$ can be written as
\begin{align}
    \nonumber \big[\mub^{(\textrm{JD})}(\x)\big]_t &= -2\Re\big[\boldsymbol{\mathsf{e}}_{{\ell_{1},\ldots,\ell_{K}}}^\herm\f_t(\x)\big] + \|\boldsymbol{{\mathsf{e}}}_{t}\|^2 \\
    \label{eq:mu xi} & \phantom{=} \ -\|\boldsymbol{{\mathsf{e}}}_{\ell_{1},\ldots,\ell_{K}}\|^2, \\
    \nonumber \big[\boldsymbol{\Sigma}^{(\textrm{JD})}(\x))\big]_{t,t'} &= 4\Big(\mathbb{E}\big[\Re\big[\hat{\x}^\herm \f_t(\x)\big]\Re\big[\hat{\x}^\herm\f_{t'}(\x)\big]\big]\\
    \label{eq:Sigma xi} & \phantom{=} \ - \Re\big[\boldsymbol{\mathsf{e}}_{{\ell_{1},\ldots,\ell_{K}}}^\herm\f_t(\x)\big]\Re\big[\boldsymbol{\mathsf{e}}_{{\ell_{1},\ldots,\ell_{K}}}^\herm\f_{t'}(\x)\big]\Big)
\end{align}
respectively, with $\f_{t}(\x) \triangleq \boldsymbol{\mathsf{e}}_{t}- \boldsymbol{\mathsf{e}}_{{\ell_{1},\ldots,\ell_{K}}}\in \mathbb{C}^K$. The term $\mathbb{E}\big[\Re\big[\hat{\x}^\herm \f_t(\x)\big]\Re\big[\f_{t'}(\x)^\herm\hat{\x}\big]\big]$ is given in \eqref{eq:Cz' JD} at the top of the page, with $\C_{\hat{\x}}^{(\mathrm{A,B})}(\x) \triangleq \mathbb{E}\big[\mathrm{A}\big[\hat{\x}\big]\mathrm{B}\big[\hat{\x}\big]|\x \big] \in \mathbb{R}^{K\times K}$ and where $\mathrm{A}[\cdot]$ and $\mathrm{B}[\cdot]$ can indicate either $\Re[\cdot]$ or $\Im[\cdot]$. The elements of $\C_{\hat{\x}}^{(\mathrm{A,B})}(\x)$ can be computed via numerical integration over the random variables $\H$, $\z$, and $\z_\mathrm{p}$, whose distributions are defined in Section~\ref{sec:Sys}. An upper bound on the probability of error of UE~$k$ for a given data symbol vector $\x$ can be derived as in \eqref{eq:Pr JD} at the top of the page, which is obtained as the complement of the positive orthant probability of $\xib^{(\textrm{JD})}$ and can be computed through numerical integration. The detailed derivations of \eqref{eq:Pr JD} are provided in Appendix~\ref{sec:App E_B}. Finally, averaging \eqref{eq:Pr JD} over all possible data symbol vectors yields an upper bound on the SER as
\setcounter{equation}{44}
\begin{align} \label{eq:SER theory JD}
\overline{\mathrm{SER}}_{k}^{\textrm{(JD)}} &= \frac{1}{L^K} \sum_{\forall \x \in \setS^{K}}\overline{\mathsf{P}}_{k}^{\textrm{(JD)}}(\x).
\end{align}
The exact probability of error of UE~$k$ for H-USD, denoted by $\mathsf{P}_{k}^{\textrm{(H-USD)}}(\x)$, is defined in \eqref{eq:SER analysis H-USD} (see Appendix~\ref{sec:App E_C}). Similar to \eqref{eq:SER theory JD}, we can obtain the exact SER for H-USD as
\begin{align} \label{eq:SER theory H-USD}
\mathrm{SER}_{k}^{\textrm{(H-USD)}} &= \frac{1}{L^K} \sum_{\forall \x \in \setS^{K}}\mathsf{P}_{k}^{\textrm{(H-USD})}(\x).
\end{align}

\section{Numerical Results and Discussion} \label{sec:NR}

In this section, we evaluate the impact of the different receivers (MRC, MMSE, and LMMD) and the data detection strategies described in Section~\ref{sec:DD_B} on the SER. We assume that the BS is equipped with a uniform linear array (ULA). Unless otherwise stated, we consider $M = 128$ antennas and $\rho = 0$~dB. All the UEs are subject to the same (normalized) pathloss, such that $\tr(\C_{\h_{k}}) = M,~\forall k = 1, \ldots, K$. The channel covariance matrices are generated based on the one-ring channel model \cite{Yin13} with angular spread of $30^{\circ}$ for each UE. In this respect, for $K = 2$ or $K = 3$ UEs, we consider angular separation of $30^{\circ}$ between the UEs. The orthogonal pilots used for the channel estimation (see Section~\ref{sec:SM_A}) are constructed as Zadoff-Chu sequences, which are widely adopted in the 4G LTE and 5G NR standards \cite{Hyd17}; unless otherwise stated, we fix $\tau = 31$. All the SER results are obtained by averaging over $4 \times 10^3$ independent channel and AWGN realizations, taking all possible data symbols into account.

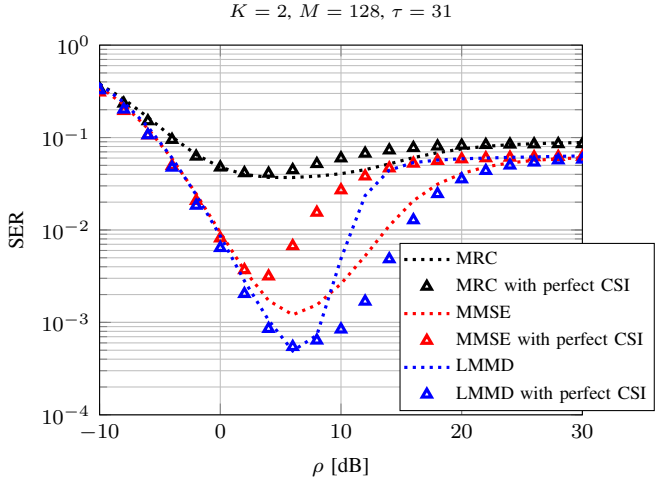
\begin{figure}[t!]
\centering
\begin{tikzpicture}

\begin{axis}[
	width=8cm,
	height=6cm,
	xmin=-10, xmax=30,
	ymin=0.0001, ymax=1,
    xlabel={$\rho$ [dB]},
    ylabel={SER},
    xlabel near ticks,
    ylabel near ticks,
    xtick={-10,0,10,20,30},
	ymode=log,
    legend style={at={(1.15,0.01)}, anchor=south east},
	legend style={font=\scriptsize, inner sep=1pt, fill opacity=0.75, draw opacity=1, text opacity=1},
    legend cell align=left,
	grid=both,
	x label style={font=\footnotesize},
	y label style={font=\footnotesize},
	ticklabel style={font=\footnotesize},
    clip marker paths=true,
    title={$K = 2$, $M = 128$, $\tau=31$},
    title style={font=\scriptsize, yshift=-1mm},
]

\addplot[very thick, black, dotted]
table [x=rho_dB, y=MRC_Mult_user_ES, col sep=comma] 
{Figures/Data/Main_SER.txt};
\addlegendentry{MRC}

\addplot[very thick, black, only marks, mark=triangle]
table [x=rho_dB, y=MRC_ideal, col sep=comma] 
{Figures/Data/Main_SER.txt};
\addlegendentry{MRC (perfect CSI)}

\addplot[very thick, red, dotted]
table [x=rho_dB, y=ZF_Mult_user_ES, col sep=comma] 
{Figures/Data/Main_SER.txt};
\addlegendentry{MMSE}


\addplot[very thick, red, only marks, mark=triangle]
table [x=rho_dB, y=ZF_ideal, col sep=comma] 
{Figures/Data/Main_SER.txt};
\addlegendentry{MMSE (perfect CSI)}

\addplot[very thick, blue, dotted]
table [x=rho_dB, y=BMMSE_MRC, col sep=comma] 
{Figures/Data/Main_SER.txt};
\addlegendentry{LMMD}

\addplot[very thick, blue, only marks, mark=triangle]
table [x=rho_dB, y=BMMSE_MRC_ideal, col sep=comma] 
{Figures/Data/Main_SER.txt};
\addlegendentry{LMMD (perfect CSI)}

\end{axis}

\end{tikzpicture} 
\caption{SER versus SNR, with JD and different receivers.}
\label{fig:SER1}
\end{figure}

Considering $K=2$, Fig.~\ref{fig:SER1} plots the SER as a function of the SNR obtained with the different receivers described in Section~\ref{sec:Ex} and~\ref{sec:Beam} and JD presented in~\ref{sec:DD_B}. We observe that all the SER curves exhibit an optimal SNR operating point:
at low SNR, the AWGN is dominant; at high SNR, the soft-estimated symbols corresponding to the data symbols with the same phase become indistinguishable (see, e.g., Fig.~\ref{fig:scatter0_b}). Between these regimes, a judicious amount of AWGN produces a beneficial variability in the 1-bit quantized signals across the $M$ antennas. This controlled scrambling improves symbol separability after combining. As a result, the performance exhibits a non-monotonic dependence on the AWGN power. This behavior is reminiscent of stochastic resonance in nonlinear systems, where an appropriate noise level can enhance detection performance. This phenomenon arises to a much lesser extent in systems employing 2- or 3-bit ADCs, since their additional quantization levels allow partial recovery of the amplitude information even at high SNR. Remarkably, MMSE provides a significant gain compared with MRC. As demonstrated in Section~\ref{sec:DD_A}, the expected values of the soft-estimated symbols with MMSE can be asymptotically obtained by simply scaling their MRC counterparts in \eqref{eq:Ex MRC}. Here, we note that Lemma~\ref{lemma1}, which leads to \eqref{eq:Ex MMSE}, does not hold with Zadoff-Chu pilots; however, \eqref{eq:as_orth} still approaches a very small value. As a result, the SER gain with MMSE stems from the reduced dispersion of the soft-estimated symbols around their expected values, rather than from the expected values themselves. 

Under imperfect CSI, MRC and MMSE achieve lower SER than with perfect CSI because the conventional receivers are not tailored to 1-bit ADCs and therefore structurally ignore the quantization distortion. In contrast, only the channel estimate within the receiver accounts for the inherent scaling ambiguity in the quantized observed signal $\mathbf{r}_{\mathrm{p}}$, which arises from the loss of amplitude information due to 1-bit quantization, as mentioned in \cite{Ngu21}. Specifically, the channel estimate in \eqref{eq:h_hat_k} incorporates $\C_{\r_{\mathrm{p}}}$, i.e., the covariance matrix of the 1-bit quantized received signal during the channel estimation, which encompasses the quantization distortion. Therefore, the conventional receivers under imperfect CSI can decorrelate the 1-bit quantized received signal $\r$ more effectively than with perfect CSI.\footnote{We note that achieving imperfect CSI by reducing the pilot length or artificially introducing pilot contamination is detrimental, as these measures do not resolve the scaling ambiguity in the channel estimate and instead only increase the estimation error. In contrast, the data detection performance may be improved through optimized pilot design, which is left for future work.} Meanwhile, the proposed LMMD exploits the Bussgang gain matrix in \eqref{eq:G(x)} to capture the proper scaling of the channel, and the covariance matrix of the 1-bit quantized signal $\C_{\r}$ used in \eqref{eq:WLMMD} allows LMMD to properly decorrelate $\r$. As a consequence, LMMD does not benefit from imperfect~CSI.

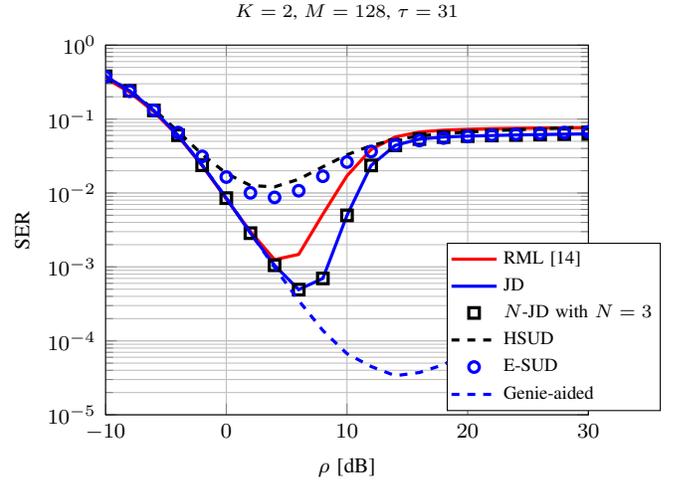
\begin{figure}[t!]
\centering
\begin{tikzpicture}

\begin{axis}[
	width=7.5cm,
	height=6cm,
	xmin=-10, xmax=30,
	ymin=0.00001, ymax=1,
    xlabel={$\rho$ [dB]},
    ylabel={SER},
    xlabel near ticks,
    ylabel near ticks,
    xtick={-10,0,10,20,30},
	ymode=log,
    legend style={at={(1.15,0.01)}, anchor=south east},
	legend style={font=\scriptsize, inner sep=1pt, fill opacity=0.75, draw opacity=1, text opacity=1},
    legend cell align=left,
	grid=both,
	x label style={font=\footnotesize},
	y label style={font=\footnotesize},
	ticklabel style={font=\footnotesize},
    clip marker paths=true,
    title={$K = 2$, $M = 128$, $\tau=31$},
    title style={font=\scriptsize, yshift=-1mm},
]

\addplot[very thick, red]
table [x=rho_dB, y=RML, col sep=comma] 
{Figures/Data/Main_SER.txt};
\addlegendentry{RML \cite{Ngu21}}

\addplot[very thick, blue, dotted]
table [x=rho_dB, y=BMMSE_MRC_ESD, col sep=comma] 
{Figures/Data/Main_SER_BMMSE.txt};
\addlegendentry{E-USD}

\addplot[very thick, blue, only marks, mark=o]
table [x=rho_dB, y=BMMSE_MRC_NJD1, col sep=comma] 
{Figures/Data/Main_SER_BMMSE.txt};
\addlegendentry{H-USD}

\addplot[very thick, green, dashed]
table [x=rho_dB, y=BMMSE_MRC_Genie, col sep=comma] 
{Figures/Data/Main_SER_BMMSE.txt};
\addlegendentry{Genie-aided}

\addplot[very thick, black]
table [x=rho_dB, y=BMMSE_MRC_JD, col sep=comma] 
{Figures/Data/Main_SER_BMMSE.txt};
\addlegendentry{JD}

\addplot[very thick, black, only marks, mark=square]
table [x=rho_dB, y=BMMSE_MRC_NJD3, col sep=comma] 
{Figures/Data/Main_SER_BMMSE.txt};
\addlegendentry{$N$-JD with $N=3$}

\end{axis}

\end{tikzpicture}
\caption{SER versus SNR, with different data detection strategies and LMMD receiver.} 
\label{fig:SER2}
\end{figure}
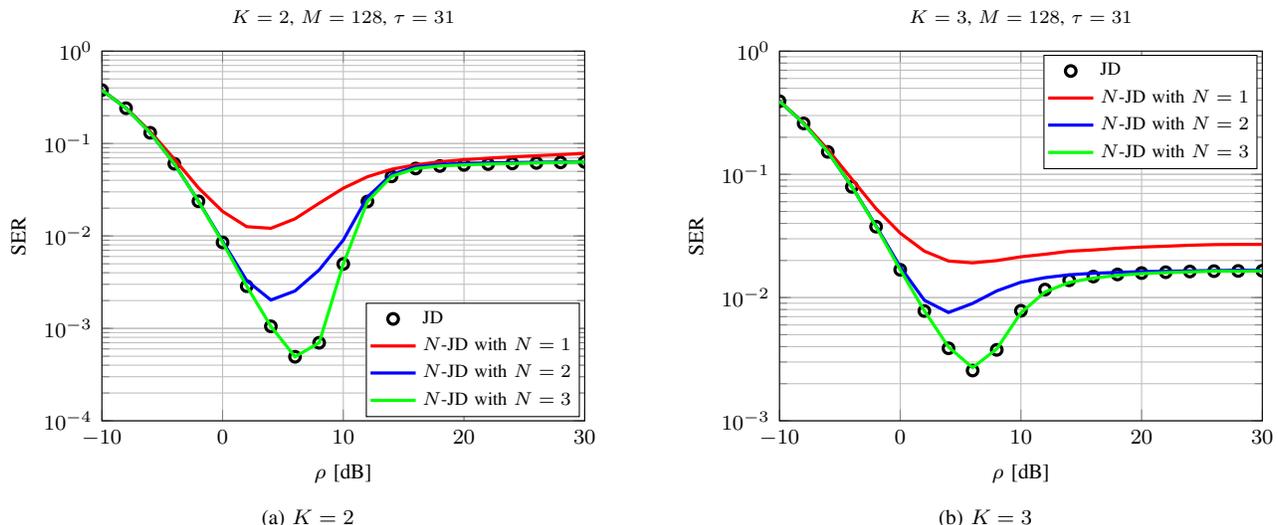
\begin{figure*}[t!]
\centering
\begin{subfigure}[c]{0.49\textwidth}
\centering
\begin{tikzpicture}

\begin{axis}[
	width=7.5cm,
	height=6cm,
	xmin=-10, xmax=30,
	ymin=0.0001, ymax=1,
    xlabel={$\rho$ [dB]},
    ylabel={SER},
    xlabel near ticks,
    ylabel near ticks,
    xtick={-10,0,10,20,30},
	ymode=log,
    legend style={at={(0.99,0.01)}, anchor=south east},
	legend style={font=\scriptsize, inner sep=1pt, fill opacity=0.75, draw opacity=1, text opacity=1},
    legend cell align=left,
	grid=both,
	x label style={font=\footnotesize},
	y label style={font=\footnotesize},
	ticklabel style={font=\footnotesize},
    clip marker paths=true,
    title={$K = 2$, $M = 128$, $\tau=31$},
    title style={font=\scriptsize, yshift=-1mm},
]


\addplot[very thick, black, only marks, mark=o]
table [x=rho_dB, y=BMMSE_MRC_JD, col sep=comma] 
{Figures/Data/Main_SER_BMMSE.txt};
\addlegendentry{JD}

\addplot[very thick, red]
table [x=rho_dB, y=BMMSE_MRC_NJD1, col sep=comma] 
{Figures/Data/Main_SER_BMMSE.txt};
\addlegendentry{$N$-JD with $N=1$}

\addplot[very thick, blue]
table [x=rho_dB, y=BMMSE_MRC_NJD2, col sep=comma] 
{Figures/Data/Main_SER_BMMSE.txt};
\addlegendentry{$N$-JD with $N=2$ }

\addplot[very thick, green]
table [x=rho_dB, y=BMMSE_MRC_NJD3, col sep=comma] 
{Figures/Data/Main_SER_BMMSE.txt};
\addlegendentry{$N$-JD with $N=3$}
\end{axis}

\end{tikzpicture}
\vspace{-1mm}
\caption{$K = 2$}
\label{fig:sphere_2UE}
\end{subfigure}
\begin{subfigure}[c]{0.49\textwidth}
\centering
\begin{tikzpicture}

\begin{axis}[
	width=7.5cm,
	height=6cm,
	xmin=-10, xmax=30,
	ymin=0.001, ymax=1,
    xlabel={$\rho$ [dB]},
    ylabel={SER},
    xlabel near ticks,
    ylabel near ticks,
    xtick={-10,0,10,20,30},
	ymode=log,
    legend style={at={(0.99,0.99)}, anchor=north east},
	legend style={font=\scriptsize, inner sep=1pt, fill opacity=0.75, draw opacity=1, text opacity=1},
    legend cell align=left,
	grid=both,
	x label style={font=\footnotesize},
	y label style={font=\footnotesize},
	ticklabel style={font=\footnotesize},
    clip marker paths=true,
    title={$K = 3$, $M = 128$, $\tau=31$},
 title style={font=\scriptsize, yshift=-1mm},
]


\addplot[very thick, black, only marks, mark=o]
table [x=rho_dB, y=BMMSE_MRC_JD, col sep=comma] 
{Figures/Data/SER_NJD_3UE_BMMSE.txt};
\addlegendentry{JD}

\addplot[very thick, red]
table [x=rho_dB, y=BMMSE_MRC_NJD1, col sep=comma] 
{Figures/Data/SER_NJD_3UE_BMMSE.txt};
\addlegendentry{$N$-JD with $N=1$}

\addplot[very thick, blue]
table [x=rho_dB, y=BMMSE_MRC_NJD2, col sep=comma] 
{Figures/Data/SER_NJD_3UE_BMMSE.txt};
\addlegendentry{$N$-JD with $N=2$}

\addplot[very thick, green]
table [x=rho_dB, y=BMMSE_MRC_NJD3, col sep=comma] 
{Figures/Data/SER_NJD_3UE_BMMSE.txt};
\addlegendentry{$N$-JD with $N=3$}

\end{axis}

\end{tikzpicture}
\vspace{-1mm}
\caption{$K = 3$}
\label{fig:sphere_3UE}
\end{subfigure}
\vspace{3mm}
\caption{SER versus SNR, with JD and $N$-JD (both with LMMD receiver).} \label{fig:sphere}
\end{figure*}

Under perfect CSI, LMMD significantly outperforms all the other receivers, as it minimizes the mean dispersion of the soft-estimated symbols around $\e(\x)$ in \eqref{eq:e(x)}. However, under imperfect CSI, LMMD is outperformed by MMSE as $\rho$ grows. In fact, while the criterion for the LMMD receiver design is to minimize the mean dispersion, it is not optimal for each individual expected value given in \eqref{eq:Ex MRC}; achieving that would require a symbol-based receiver, which is clearly impractical. In other words, at high SNR, LMMD forces groups of soft-estimated symbols to have minimum dispersion around the expected values that correspond to data symbols with the same phase but different amplitudes, since these expected values become increasingly indistinguishable, as discussed in Section~\ref{sec:Ex}. In contrast, MMSE treats all the soft-estimated symbols uniformly across the entire SNR range. Nevertheless, Fig.~\ref{fig:SER1} shows that LMMD significantly outperforms MMSE around $\rho = 6$~dB, which is the optimal SNR operating point. In the rest of this section, unless otherwise stated, we use the LMMD receiver.

Considering $K=2$ UEs, Fig.~\ref{fig:SER2} plots the SER as a function of the SNR obtained with all the data detection strategies described in Section~\ref{sec:DD}. For comparison, we also include the RML method with NN search from \cite{Ngu21}, where the corresponding formulation is summarized for convenience in Appendix~\ref{sec:App D}. In order to achieve the maximum performance of RML, we consider a full search over all possible data symbol vectors. We observe that E-USD outperforms H-USD since the latter considers the values of $\bar{\mathsf{e}}_{k,l_{k}}$ in \eqref{eq:E_kl}. Moreover, the genie-aided data detection outperforms JD, as it assumes perfect knowledge of the data symbols transmitted by the interfering UEs, as detailed in Section~\ref{sec:DD_A}. We note that this scheme is impractical and serves only to provide a lower bound on the SER. Remarkably, JD and $N$-JD with $N = 3$ provide a significant boost over E-USD. This means that taking advantage of the interdependence among the soft-estimated symbols of the interfering UEs gives a notable gain over the data detection strategies that treat each UE individually. Lastly, while RML performs similarly to the proposed JD and $N$-JD strategies at low SNR, its performance deteriorates at moderate-to-high SNR. As the SNR increases, the data symbols with the same phase result in nearly identical 1-bit quantized received signals, and the RML detection problem in \eqref{eq:RML} provides the same solution for these points. Therefore, while RML is designed to be robust against imperfect CSI, it does not account for this phenomenon and is thus inferior to JD and $N$-JD.

Considering $K = 2$ UEs, Fig.~\ref{fig:sphere_2UE} illustrates that $N$-JD exhibits the same performance with respect to JD with $N = 3$. This is because, with $K = 2$, there is significant overlap among many of the $256$ expected values of the soft-estimated symbols. In addition, since there are three different amplitude levels in the 16-QAM constellation, only $3 \times 16 = 48$ expected values can be clearly distinguished (see Fig.~\ref{fig:E1}). Fig.~\ref{fig:sphere_3UE} extends the insights of Fig.~\ref{fig:sphere_2UE} to the case of $K = 3$. Here, there are $16^3 = 4096$ different triplets of data symbols transmitted by the three UEs, each corresponding to a different expected value (see Fig.~\ref{fig:scatter_K=3}). Remarkably, $N$-JD with $N =3$ nearly matches JD while its computational complexity is only $3 \times$ higher than for $K = 2$. As shown in Figs.~\ref{fig:sphere_2UE} and~\ref{fig:sphere_3UE}, the SER degrades significantly at very high SNR, where the amplitude information in the 16-QAM data symbols cannot be recovered. Moreover, the channel estimation error saturates at high SNR for a fixed value of $\tau$, as detailed in \cite{Atz22}. To address this issue, one should either transmit PSK data symbols instead of QAM, or judiciously apply UE power control \cite{Rav23}. Additionally, one can increase the number of UEs to make the signal more robust against the quantization distortion, albeit at the cost of higher interference and computational complexity. The impact of increasing the number of UEs can be seen by comparing Figs.~\ref{fig:sphere_2UE} and~\ref{fig:sphere_3UE}, where the SER for the case of $K = 3$ UEs is about $4 \times$ lower than for $K = 2$ around $\rho = 30$~dB.

\begin{figure}[t!]
\centering
\begin{tikzpicture}

\begin{axis}[
	width=7.5cm,
	height=5cm,
	xmin=8, xmax=256,
	ymin=1e-6, ymax=1,
    xlabel={$M$},
    ylabel={Minimum SER over $\rho$},
    xlabel near ticks,
	ylabel near ticks,
    xtick={8,16,32,64,128,256},
    ytick={0.000001,0.00001,0.0001,0.001,0.01,0.1,1},
    yticklabels={$10^{-6}$,$10^{-5}$,$10^{-4}$,$10^{-3}$,$10^{-2}$,$10^{-1}$,$10^{0}$},
	log ticks with fixed point,
	ymode=log,
   legend pos=north east,
    legend style={at={(0.01,0.01)}, anchor=south west},
	legend style={font=\scriptsize, inner sep=1pt, fill opacity=0.75, draw opacity=1, text opacity=1},
    legend cell align=left,
	grid=both,
	x label style={font=\footnotesize},
	y label style={font=\footnotesize},
	ticklabel style={font=\footnotesize},
    clip marker paths=true,
    title={$K = 2$, $\tau=31$},
    title style={font=\scriptsize, yshift=-1mm},
    xmode=log,
	log basis x={2},
]


\addplot[line width=1pt, black]
table [x=num_BS, y=SER1, col sep=comma] 
{Figures/Data/MinSER2_BMMSE.txt};
\addlegendentry{$N$-JD with $N=1$}


\addplot[line width=1pt, red]
table [x=num_BS, y=SER3, col sep=comma] 
{Figures/Data/MinSER2_BMMSE.txt};
\addlegendentry{$N$-JD with $N=3$}

\end{axis}

\end{tikzpicture}
\caption{Minimum SER versus number of BS antennas, with $N$-JD and LMMD receiver.}\label{fig:min_SER num_BS}
\end{figure}
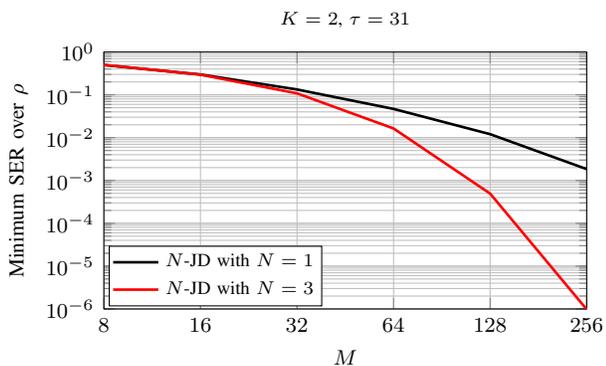

Considering $K = 2$ UEs, Fig.~\ref{fig:min_SER num_BS} illustrates the impact of the number of BS antennas $M$ on the minimum achievable SER over the SNR $\rho$. We observe that the minimum SER decreases monotonically with increasing $M$. This improvement stems from the higher granularity in the antenna domain provided by larger arrays, which enhances the scrambling of 1-bit quantized signals and reduces the dispersion of the soft-estimated symbols around their expected values. Conversely, when $M$ is small, the benefit of JD becomes negligible because severe quantization distortion makes the candidate points (within each group of the expected values of the soft-estimated symbols) indistinguishable. We emphasize that, since the ADC power consumption grows exponentially with the resolution, a massive array equipped with low-resolution or 1-bit ADCs achieves a higher array gain than a system with a limited number of antennas and high-resolution ADCs under the same power budget at the BS.

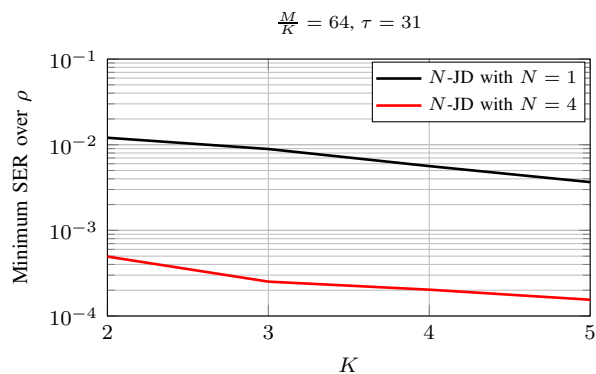
\begin{figure}[t!]
    \centering
    \begin{tikzpicture}

\begin{axis}[
	width=7.5cm,
	height=5cm,
	xmin=2, xmax=6,
	ymin=1e-3, ymax=1e-1,
    xlabel={$K$},
    ylabel={Minimum SER over $\rho$},
    xlabel near ticks,
    ylabel near ticks,
    xtick={2,3,4,5,6},
	ymode=log,
    legend style={at={(0.99,0.01)}, anchor=south east},
	legend style={font=\scriptsize, inner sep=1pt, fill opacity=0.75, draw opacity=1, text opacity=1},
    legend cell align=left,
	grid=both,
	x label style={font=\footnotesize},
	y label style={font=\footnotesize},
	ticklabel style={font=\footnotesize},
    clip marker paths=true,
    title={$\frac{M}{K} = 32$, $\tau=31$},
    title style={font=\scriptsize, yshift=-1mm},
]
\addplot[very thick, black]
table [x=num_UE, y=min_SERN1, col sep=comma] 
{Figures/Data/minSER_numUE_LMMD_MMSE_load32_2.txt};
\addlegendentry{LMMD $N$-JD with $N=1$}

\addplot[very thick, red]
table [x=num_UE, y=min_SERN4, col sep=comma] 
{Figures/Data/minSER_numUE_LMMD_MMSE_load32_2.txt};
\addlegendentry{LMMD $N$-JD with $N=4$}



\end{axis}

\end{tikzpicture}
    \caption{Minimum SER versus number of UEs, with $N$-JD and LMMD receiver.}\label{fig:fixed load}
\end{figure}

Fig.~\ref{fig:fixed load} plots the minimum SER over the SNR $\rho$ against the number of UEs for $N$-JD with a fixed load of $\frac{M}{K}=32$. In this figure, the UEs are equally spaced in the angular sector spanning from $45^{\circ}$ to $135^{\circ}$ with respect to the ULA's endfire direction. For all the values of $K$, the target UE is placed at $45^{\circ}$ and the angular spread is set to $30^{\circ}$ for all the UEs. We observe that $N$-JD with $N=4$ and $N=1$ exhibit an approximately constant SER gap. Up to $K = 5$, the SER decreases due to the improved spatial resolution offered by larger arrays, whose size grows proportionally with $K$. However, the SER increases for $K = 6$ as a result of the stronger interference caused by the growing overlap among the UEs' angular supports.

Lastly, for $K = 2$~UEs, Fig.~\ref{fig:theory & sim} validates the analytical SER expressions in Section~\ref{sec:DD_D}, derived for MRC under i.i.d. Rayleigh fading for tractability. We first observe that the analytical SER curves closely match the simulation results. Furthermore, the curves corresponding to MRC under i.i.d. Rayleigh fading serve as an upper bound on those obtained with MMSE under correlated Rayleigh fading.

\begin{figure}[t!]
    \centering
    \begin{tikzpicture}

\begin{axis}[
	width=7.5cm,
	height=6cm,
	xmin=-10, xmax=30,
	ymin=0.001, ymax=1,
    xlabel= {$\rho$~[dB]},
    ylabel={SER},
    xlabel near ticks,
    ylabel near ticks,
    xtick={-10,0,10,20,30},
	ymode=log,
    legend style={at={(0.01,0.99)}, anchor=north west},
	legend style={font=\scriptsize, inner sep=1pt, fill opacity=0.75, draw opacity=1, text opacity=1},
    legend cell align=left,
	grid=both,
	x label style={font=\footnotesize},
	y label style={font=\footnotesize},
	ticklabel style={font=\footnotesize},
    clip marker paths=true,
    title={$K = 2$, $M = 128$, $\tau=31$},
    title style={font=\scriptsize, yshift=-1mm},
]

\addplot[very thick, black]
table [x=rhodB, y=SER_the, col sep=comma] 
{Figures/Data/SER_analysis_HSUD.txt};
\addlegendentry{MRC and i.i.d. Rayleigh fading (\eqref{eq:SER theory JD} and \eqref{eq:SER theory H-USD})}

\addplot[very thick, black, forget plot]
table [x=rhodB, y=SER_the, col sep=comma] 
{Figures/Data/SER_analysis_JD.txt};

\addplot[very thick, black, only marks, mark=star]
table [x=rhodB, y=SER_MRC, col sep=comma] 
{Figures/Data/SER_analysis_HSUD.txt};
\addlegendentry{MRC and i.i.d. Rayleigh fading (simulations)}

\addplot[very thick, black, only marks, mark=star, forget plot]
table [x=rhodB, y=SER_MRC, col sep=comma] 
{Figures/Data/SER_analysis_JD.txt};

\addplot[very thick, red, dotted]
table [x=rho_dB, y=Full_multiUE, col sep=comma] 
{Figures/Data/SER_NP_JD.txt};
\addlegendentry{MMSE and correlated Rayleigh fading (simulations)}

\addplot[very thick, red, dotted, forget plot]
table [x=rho_dB, y=Sphere_N1, col sep=comma] 
{Figures/Data/SER_NP_JD.txt};

\draw[] (10,0.05) ellipse [x radius=0.1cm, y radius=0.3cm];
\node[font=\scriptsize] at (7, 0.09) {H-USD};

\draw[] (12,0.01) ellipse [x radius=0.3cm, y radius=0.6cm];
\node[font=\scriptsize] at (14, 0.0035) {JD};
\end{axis}

\end{tikzpicture}
    \caption{Analytical and simulated SER versus SNR, with JD and H-USD and different receivers.}
 \label{fig:theory & sim}
\end{figure}

\section{Conclusions} \label{sec:Concl}

Considering a multi-UE setting with correlated Rayleigh fading, we investigated the uplink data detection in a single-cell massive MIMO system with 1-bit ADCs.  We first characterized the expected values of the soft estimated symbols, which effectively capture the distortion in the transmit constellation due to 1-bit quantization during both the channel estimation and the uplink data transmission. In our analysis, we considered conventional receivers such as MRC, ZF, and MMSE. Additionally, we designed the \textit{linear minimum mean dispersion} (LMMD) receiver tailored for data detection with 1-bit ADCs, which exploits the expected values of the soft-estimated symbols. Then, we proposed a joint data detection (JD) strategy that exploits the interdependence among the soft-estimated symbols of the interfering UEs, along with its low-complexity variant. Numerical results examining the SER showed that the proposed LMMD receiver significantly outperforms all the conventional receivers. Lastly, the proposed JD and its low-complexity variant provided a significant boost in comparison with UE-specific data detection. Future work will analyze the impact of pilot design and propose data detection strategies (potentially data-driven) tailored for large numbers of UEs.

\appendices 

\section{Proof of Theorem~\ref{thm:main}} \label{sec:App A}

\begin{figure*}[t!]
\setcounter{equation}{56}
\begin{align} \label{eq:Er_m rp_n,u (2)}
    \nonumber \mathbb{E}[r_m r_{\mathrm{p},n,u}^*] &= (\rho K + 1)\bigg[ \Omega\bigg(\frac{\rho \sum_{k=1}^{K}{\sum_{i=1}^{M} \Re\big[[\C_{\h_k}^{\frac{1}{2}}]_{m,i} [\C_{\h_k}^{\frac{1}{2}}] _{n,i}^*  x_{k} P_{u.k}\big] }} {\big(\rho\sum_{k=1}^{K} \sum_{i=1}^{M} \big|[\C_{\h_{k}}^{\frac{1}{2}}] _{m,i} x_{k}\big|^2 + 1\big)^{\frac{1}{2}} \ \big(\rho \sum_{k=1}^{K} \sum_{i=1}^{M} \big|[\C_{\h_{k}}^{\frac{1}{2}}] _{n,i}\big|^2 + 1\big)^{\frac{1}{2}}}\bigg) \\
    & \phantom{=} \ + j \, \Omega\bigg(\frac{\rho \sum_{k=1}^{K}{\sum_{i=1}^{M} \Im\big[[\C_{\h_k}^{\frac{1}{2}}] _{m,i} [\C_{\h_k}^{\frac{1}{2}}] _{n,i}^*  x_{k} P_{u.k}\big] }} {\big(\rho\sum_{k=1}^{K} \sum_{i=1}^{M} \big|[\C_{\h_{k}}^{\frac{1}{2}}] _{m,i} x_{k}\big|^2 + 1\big)^{\frac{1}{2}} \ \big(\rho \sum_{k=1}^{K} \sum_{i=1}^{M} \big|[\C_{\h_{k}}^{\frac{1}{2}}] _{n,i}\big|^2 + 1\big)^{\frac{1}{2}}}\bigg) \bigg]
\end{align}
\hrulefill
\begin{align} \label{eq:a_m A_{n,u} (1)}
    \mathbb{E}[a_m A_{n,u}] &= \Omega\bigg(\frac{\rho \sum_{k=1}^{K}{\sum_{i=1}^{M} \Re\big[[\C_{\h_k}^{\frac{1}{2}}]_{m,i} [\C_{\h_k}^{\frac{1}{2}}]_{n,i}^* x_{k} P_{u.k}\big]}} {\big(\rho\sum_{k=1}^{K} \sum_{i=1}^{M} \big|[\C_{\h_{k}}^{\frac{1}{2}}] _{m,i} x_{k}\big|^2 + 1\big)^{\frac{1}{2}} \ \big(\rho \sum_{k=1}^{K} \sum_{i=1}^{M} \big|[\C_{\h_{k}}^{\frac{1}{2}}] _{n,i}\big|^2 + 1\big)^{\frac{1}{2}}}\bigg)
\end{align}
\hrulefill
\begin{align}
    \nonumber \zetab &= \big[\Re[\bar{h}_{1,1}], \ldots, \Re[\bar{h}_{M,1}], \ldots, \Re[\bar{h}_{1,K}], \ldots,\Re[\bar{h}_{M,K}], \Im[\bar{h}_{1,1}], \ldots, \Im[\bar{h}_{M,1}] ,\ldots, \\
    & \phantom{=} \ \Im[\bar{h}_{1,K}], \ldots, \Im[\bar{h}_{M,K}], \Re[z_{\mathrm{p}_{n,u}}], \Re[z_m] \big]^\tran \sim \mathcal{N} \bigg( \0_{(2MK+2)},\frac{1}{2}\I_{(2MK+2)} \bigg), \label{zeta} \\
    \nonumber \a_{1}^{(m)} &= \sqrt\rho\Big[\Re\big[[\C_{\h_{1}}^{\frac{1}{2}}] _{m,1}x_{1}\big], \ldots, \Re\big[[\C_{\h_{1}}^{\frac{1}{2}}] _{m,M}x_{1}\big], \ldots, \Re\big[[\C_{\h_{K}}^{\frac{1}{2}}] _{m,1}x_{K}\big], \ldots, \Re\big[[\C_{\h_{K}}^{\frac{1}{2}}]_{m,M}x_{K}\big], \\
    \nonumber & \phantom{=} \ -\Im\big[[\C_{\h_{1}}^{\frac{1}{2}}] _{m,1}x_{1}\big], \ldots, -\Im\big[[\C_{\h_{1}}^{\frac{1}{2}}] _{m,M}x_{1}\big], \ldots, -\Im\big[[\C_{\h_{K}}^{\frac{1}{2}}] _{m,1}x_{K}\big], \ldots, -\Im\big[[\C_{\h_{K}}^{\frac{1}{2}}] _{m,M}x_{K}\big],0,1 \Big]^\tran \\
    & \in \mathbb{R}^{(2MK+2)}, \label{a_1} \\
    \nonumber \a_{2}^{(n,u)} &= \sqrt\rho\Big[\Re\big[[\C_{\h_{1}}^{\frac{1}{2}}] _{n,1}P_{u,1}^*\big], \ldots, \Re\big[[\C_{\h_{1}}^{\frac{1}{2}}] _{n,M}P_{u,1}^*\big], \ldots, \Re\big[[\C_{\h_{K}}^{\frac{1}{2}}] _{n,1}P_{u,K}^*\big], \ldots, \Re\big[[\C_{\h_{K}}^{\frac{1}{2}}] _{n,M}P_{u,K}^*\big], \\
    \nonumber & \phantom{=} \ -\Im\big[[\C_{\h_{1}}^{\frac{1}{2}}] _{n,1}P_{u,1}^*\big] , \ldots, -\Im\big[[\C_{\h_{1}}^{\frac{1}{2}}] _{n,M}P_{u,1}^*\big], \ldots, -\Im\big[[\C_{\h_{K}}^{\frac{1}{2}}] _{n,1}P_{u,K}^*\big], \ldots, -\Im\big[[\C_{\h_{K}}^{\frac{1}{2}}] _{n,M}P_{u,K}^*\big]
    ,1,0 \Big]^\tran \\
    & \in \mathbb{R}^{(2MK+2)} \label{a_2}
\end{align}

\vspace{-4mm}

\hrulefill
\begin{align} \label{eq:a_m A_{n,u} (2)}
    \nonumber&\mathbb E \bigg[ \sgn \bigg( \sqrt{\rho} \sum_{k=1}^{K} \sum_{i = 1}^{M}\big(\Re[\bar{h}_{i,k}] \Re\big[[\C_{\h_{k}}^{\frac{1}{2}}] _{m,i}x_{k}\big] - \Im[\bar{h}_{i,k}] \Im\big[[\C_{\h_{k}}^{\frac{1}{2}}] _{m,i}x_{k}\big]\big) + \Re[z_{m}] \bigg) \nonumber\\
    & \times\sgn \bigg( \sqrt{\rho} \sum_{k=1}^{K} \sum_{i = 1}^{M}\big(\Re[\bar{h}_{i,k}] \Re\big[[\C_{\h_{k}}^{\frac{1}{2}}] _{n,i}P_{u,k}^*\big] - \Im[\bar{h}_{i,k}] \Im\big[[\C_{\h_{k}}^{\frac{1}{2}}] _{n,i}P_{u,k}^*\big]\big) + \Re[z_{\mathrm{p}_{n,u}}]  \bigg) \bigg] \nonumber\\
    & = \Omega\left(\frac{\rho \sum_{k=1}^{K}{\sum_{i=1}^{M} \big(\Re\big[[\C_{\h_k}^{\frac{1}{2}}]_{m,i} x_{k}\big] \Re\big[[\C_{\h_k}^{\frac{1}{2}}]_{n,i} P_{u,k}^*\big] + \Im\big[[\C_{\h_k}^{\frac{1}{2}}]_{m,i} x_{k}\big] \Im\big[[\C_{\h_k}^{\frac{1}{2}}]_{n,i} P_{u,k}^*\big]\big)}} {\Big(\rho\sum\limits_{k=1}^{K} \sum\limits_{i=1}^{M}\big(\Re\big[[\C_{\h_k}^{\frac{1}{2}}]_{n,i} P_{u,k}^*\big]^2 \! + \! \Im\big[[\C_{\h_k}^{\frac{1}{2}}]_{n,i} P_{u,k}^*\big]^2 \big) \! + \! 1\Big)^{\frac{1}{2}} \Big(\rho\sum\limits_{k=1}^{K} \sum\limits_{i=1}^{M}\big(\Re\big[[\C_{\h_{k}}^{\frac{1}{2}}] _{m,i} x_{k}\big]^2 \! + \! \Im\big[[\C_{\h_{k}}^{\frac{1}{2}}] _{m,i} x_{k}\big]^2 \big) \! + \! 1\Big)^{\frac{1}{2}}}\right)
\end{align}
\hrulefill
\vspace{-4mm}
\end{figure*}

Considering \eqref{eq:h_hat_k} and \eqref{eq:r}, for a given data symbol vector $\x$, the expected value of the soft-estimated symbol of UE~$k$ obtained with the MRC receiver is $\mathsf{e}_{k}^{\textrm{(MRC)}} = \mathbb E [\hat{\h}_{k}^\herm\r]$ given by \eqref{eq:Ex MRC}.
Next, we derive closed-form expressions of $\C_{\r\r_\mathrm{p}}$ and $\C_{\r_\mathrm{p}}$ defined in Sections~\ref{sec:Ex_A} and~\ref{sec:SM_A}, respectively. To this end, let $\bar{\h}_k \triangleq \C_{\h_{k}}^{-\frac{1}{2}} \h_k$, and let $\bar{h}_{i,k}$ and $h_{i,k}$ denote the $i$th element of $\bar{\h}_k$ and $\h_k$, respectively, with
\setcounter{equation}{46}
\begin{align}
    h_{m,k} & = \sum_{i = 1}^{M} [\C_{\h_{k}}^{\frac{1}{2}}]_{m,i}\bar{h}_{i,k}, \\
    \nonumber\Re[h_{m,k}] &= \sum_{i =1}^{M} \Big(\Re\big[[\C_{\h_{k}}^{\frac{1}{2}}] _{m,i}\big] \Re[\bar{h}_{i,k}] \\& \phantom{=} \ -\Im\big[[\C_{\h_{k}}^{\frac{1}{2}}]_{m,i}\big] \Im[\bar{h}_{i,k}]\Big), \\
    \nonumber\Im[h_{m,k}] &= \sum_{i = 1}^{M} \Big(\Re\big[[\C_{\h_{k}}^{\frac{1}{2}}] _{m,i}\big] \Im[\bar{h}_{i,k}] \\&\phantom{=} \ 
     + \Im\big[[\C_{\h_{k}}^{\frac{1}{2}}] _{m,i}] \Re[\bar{h}_{i,k}\big]\Big).
\end{align}
Moreover, we introduce the following preliminary definitions:
\begin{align}
\label{Amu} 
    \nonumber A_{m,u} &\triangleq  \sgn \bigg( \sqrt{\rho} \sum_{k=1}^{K} \sum_{i = 1}^{M}\big(\Re[\bar{h}_{i,k}] \Re\big[[\C_{\h_{k}}^{\frac{1}{2}}] _{m,i}P_{u.k}^*\big] \\&\phantom{=} \ - \Im[\bar{h}_{i,k}] \Im\big[[\C_{\h_{k}}^{\frac{1}{2}}] _{m,i}P_{u.k}^*\big]\big) + \Re[z_{\mathrm{p}_{m,u}}]  \bigg),\\
    \nonumber B_{m,u} &\triangleq  \sgn \bigg( \sqrt{\rho} \sum_{k=1}^{K} \sum_{i = 1}^{M}\big(\Re[\bar{h}_{i,k}] \Im\big[[\C_{\h_{k}}^{\frac{1}{2}}] _{m,i}P_{u.k}^*\big] \\&\phantom{=} \ + \Im[\bar{h}_{i,k}] \Re\big[[\C_{\h_{k}}^{\frac{1}{2}}] _{m,i}P_{u.k}^*\big]\big) + \Im[z_{\mathrm{p}_{m,u}}]  \bigg), \\
\label{am}
    \nonumber a_{m} &\triangleq  \sgn \bigg( \sqrt{\rho} \sum_{k=1}^{K} \sum_{i = 1}^{M}\big(\Re[\bar{h}_{i,k}] \Re\big[[\C_{\h_{k}}^{\frac{1}{2}}] _{m,i}x_{k}\big] \\&\phantom{=} \ - \Im[\bar{h}_{i,k}] \Im\big[[\C_{\h_{k}}^{\frac{1}{2}}] _{m,i}x_{k}\big]\big) + \Re[z_{m}]  \bigg),
\end{align}
\begin{align}
\label{bm}
    \nonumber b_{m} &\triangleq  \sgn \bigg( \sqrt{\rho} \sum_{k=1}^{K} \sum_{i = 1}^{M}\big(\Re[\bar{h}_{i,k}] \Im\big[[\C_{\h_{k}}^{\frac{1}{2}}] _{m,i}x_{k}\big] \\&\phantom{=} \ + 
    \Im[\bar{h}_{i,k}] \Re\big[[\C_{\h_{k}}^{\frac{1}{2}}] _{m,i}x_{k}\big]\big) + \Im[z_{m}]  \bigg).
\end{align}
The following proposition will be used in the derivations. The proof is based on \cite[App.~II]{Atz22} and is thus omitted.

\begin{proposition} \label{prop1}
Let  $\zetab \sim
\setN(\0_{N},\,\gamma\I_N) $. For $\a_1, \a_2 \in \ \mathbb{R}^{N}$, we have
\begin{align}
\label{Prepos}
    \mathbb{E} \big[ \sgn(\a_1 ^\tran \zetab) \sgn(\a_2 ^\tran \zetab) \big] = \Omega \bigg(\frac{\a_1 ^\tran \a_2}{\|\a_1\| \, \|\a_2\|} \bigg).
\end{align}
\end{proposition}

\smallskip

We can write the $(m, (u-1)M + n)$th element of $\C_{\r \r_\mathrm{p}}$ as 
\begin{align}
    \mathbb{E}[r_m r_{\mathrm{p},n,u}^*] &= \frac{\rho K + 1}{2} \Big( \mathbb{E}\big[(a_m +j \, b_m) (A_{n,u} -j \, B_{n,u})\big] \Big) \\ & = (\rho K + 1) \big(\mathbb{E}[a_m A_{n,u}] +  j \, \mathbb{E}[b_m A_{n,u}] \big) \label{eq:Er_m rp_n,u (1)}
\end{align}
with $\mathbb{E}[a_m A_{n,u}] = \mathbb{E}[b_m B_{n,u}]$ and $\mathbb{E}[b_m A_{n,u}]  = - \mathbb{E}[a_m B_{n,u}]$. Hence, \eqref{eq:Er_m rp_n,u (1)} is expressed as in \eqref{eq:Er_m rp_n,u (2)} at the top of the page. 
For instance, $\mathbb{E}[a_m A_{n,u}]$ in \eqref{eq:a_m A_{n,u} (1)} can be obtained by plugging \eqref{zeta}--\eqref{a_2} at the top of the page into Proposition~\eqref{prop1}, which gives \eqref{eq:a_m A_{n,u} (2)} at the top of the page. Finally, the expression in \eqref{eq:Crrp} readily follows. Similarly, we can write the ($(u-1)M+m,(v-1)M+n$)th element of $\mathbf{C}_{\r_\mathrm{p}}$ as
\setcounter{equation}{62}
\begin{align}
    \mathbb{E}[r_{\mathrm{p},m,u} r_{\mathrm{p},n,v}^*]
    & = \frac{\rho K + 1}{2} \big(\mathbb{E}[A_{m,u} A_{n,v}] + \mathbb{E}[B_{m,u} B_{n,v}] 
   \nonumber \\& \phantom{=} \ -j \, \mathbb{E}[A_{m,u} B_{n,v}] +  j \, \mathbb{E}[B_{m,u} A_{n,v}] \big) \label{eq:Erp_mu rp_nv (1)}
\end{align}
with $\mathbb{E}[A_{m,u} A_{n,v}] = \mathbb{E}[B_{m,u} B_{n,v}] $ and $\mathbb{E}[A_{m,u} B_{n,v}] = -\mathbb{E}[B_{m,u} A_{n,v}]$. The rest of the derivations follow similar steps as in \eqref{eq:Er_m rp_n,u (2)}--\eqref{eq:a_m A_{n,u} (2)}, which finally yield the expression in \eqref{eq:Crp}.

\begin{figure*}[t!]
\setcounter{equation}{77}
\begin{align}\label{eq:Cyr|x (1) (App)}
    \nonumber[\C_{\y\r|\x}]_{n,m} &= \sqrt{\frac{\rho K +1}{2}} \Big( \mathbb{E} \big[ \Re[y_n] \sgn \big( \Re[y_m] \big) \big] + \mathbb{E} \big[ \Im[y_n] \sgn \big( \Im[y_m] \big) \big] \\
    & \phantom{=} \ - j \, \mathbb{E} \big[ \Re[y_n] \sgn \big( \Im[y_m] \big) \big] + j \, \mathbb{E} \big[ \Im[y_n] \sgn \big( \Re[y_m] \big) \big] \Big) \\
    & = \begin{cases}
    \sqrt{\frac{\rho K +1}{2}} \Big( \sqrt{\frac{1}{\pi}} \big( \exp \big( -\Re[\zeta_m]^2 \big) + \exp \big( -\Im[\zeta_m]^2 \big) \big) + \Re \big[ \zeta_n \erf_{\textrm{c}}(\zeta_m^*) \big] +j \, \Im \big[ \zeta_n \erf_{\textrm{c}}(\zeta_m^*) \big] \Big) & \quad \! \! \textrm{if}~m=n, \\
    \sqrt{\frac{\rho K +1}{2}} \Big( \Re \big[ \zeta_n \erf_{\textrm{c}}(\zeta_m^*) \big] +j \, \Im \big[ \zeta_n \erf_{\textrm{c}}(\zeta_m^*) \big] \Big) & \quad \! \! \textrm{otherwise}
    \end{cases}\label{eq:Cyr|x (2) (App)}
\end{align}
\hrulefill
\begin{align} \label{eq:Ey_n sgn(y_m) (App)}
    \mathbb{E} \big[ \Re[y_n] \, \sgn \big( \Re[y_m] \big) \big] & =
    \begin{cases}
    \sqrt{\frac{1}{\pi}} \exp \big( -\Re[\zeta_m]^2 \big) + \Re[\zeta_n] \erf \big( \Re[\zeta_m] \big) & \quad \textrm{if}~m=n, \\
    \Re[\zeta_n] \erf \big( \Re[\zeta_m] \big) &\quad \textrm{otherwise} 
    \end{cases}
\end{align}
\hrulefill
\vspace{-4mm}
\end{figure*}

\section{Proof of Lemma~\ref{lemma1}} \label{sec:App B}

The proof follows similar steps as the proof of the asymptotically favorable propagation of massive MIMO channels in \cite[Sec.~2.5.2]{massivemimobook}. For a pair of channel estimates $\hat{\h}_k$ and $\hat{\h}_{k'}$, with $k\neq k'$, asymptotic orthogonality holds if \eqref{eq:as_orth} is satisfied. According to the definitions in Section~\ref{sec:SM_A}, we use the Bussgang decomposition  to rewrite $\r_\mathrm{p}$ in \eqref{rp} as \cite{Li17}
\setcounter{equation}{63}
\begin{align}
    \r_\mathrm{p} = \A_\mathrm{p} \y_\mathrm{p} + \q_\mathrm{p} = \sqrt{\rho}\A_\mathrm{p} \bar{\P}^{*}\h
    + \A_\mathrm{p} \z_\mathrm{p} + \q_\mathrm{p}
\end{align}
where $\q_\mathrm{p} \in \mathbb{C}^{M\tau}$ is the quantization distortion vector that is uncorrelated with $\y_\mathrm{p}$. Let $\n \triangleq \A_\mathrm{p} \z_\mathrm{p} + \q_\mathrm{p} \in \mathbb{C}^{M\tau}$ denote the effective noise with zero mean and covariance matrix
\begin{align} \label{eq:Sigma_n}
    \hspace{-1mm} \C_{\n} \triangleq \Exp[\n \n^{\herm}] = \C_{\r_\mathrm{p}} -\rho \A_{\mathrm{p}} \bar{\P}^{*}\C_{\h}\bar{\P}^\tran \A_{\mathrm{p}} \in \mathbb{C}^{M\tau \times M\tau}.
\end{align}
In addition, let $\X_k \triangleq \C_{\h_k}\bar{\P}_k^\tran \A_{\mathrm{p}}\C_{\r_{\mathrm{p}}}^{-1} \in \mathbb{C}^{M\times M\tau}$. From \eqref{eq:h_hat_k} and \eqref{eq:Eh_hat}, we can write
\begin{align}
    \hat{\h}_k^\herm \hat{\h}_{k'} &= \rho \r_{\mathrm{p}} ^\herm  \X_{k}^\herm \X_{k'}\r_{\mathrm{p}} \\& = \label{eq:h'h (2)}\rho^2\bar{\h}^\herm \C_{\h}^{\frac{1}{2}}\bar{\P}^\tran \A_\mathrm{p} \X_k^\herm \X_{k'}\A_\mathrm{p} \bar{\P}^{*}\C_{\h}^{\frac{1}{2}}\bar{\h} \nonumber\\& \phantom{=} \ +  \rho^{\frac{3}{2}}\bar{\h}^\herm \C_{\h}^{\frac{1}{2}}\bar{\P}^\tran \A_\mathrm{p} \X_k^\herm \X_{k'}\C_{\n}^{\frac{1}{2}}\bar{\n} \nonumber\\& \phantom{=} \ +  \rho^{\frac{3}{2}}\bar{\n}^\herm \C_{\n}^{\frac{1}{2}}  \X_k^\herm \X_{k'}\A_\mathrm{p} \bar{\P}^{*}\C_{\h}^{\frac{1}{2}}\bar{\h} \nonumber\\& \phantom{=} \ +  \bar{\n}^\herm \C_{\n}^{\frac{1}{2}}  \X_k^\herm \X_{k'}\C_{\n}^{\frac{1}{2}}\bar{\n} 
\end{align}
with $\bar{\h} \triangleq \C_{\h}^{-\frac{1}{2}}\h$ and $\bar{\n} \triangleq \C_{\n}^{-\frac{1}{2}}\n$, where the latter can be modeled as a $\mathcal{CN}(\0_{M \tau},\I_{M\tau})$ random vector as in \cite{Ngu21}. From \eqref{eq:Eh_hat}, we have $\mathbb E[\| \hat{\h}_k\|^2] = \rho \tr(\X_k \C_{\r_{\mathrm{p}}} \X_k^\herm)$ and, for a given $M$ and average eigenvalue $\beta_\mathrm{av}$ of $\X_k \C_{\r_{\mathrm{p}}} \X_k^\herm$, it follows that $\mathbb E[\| \hat{\h}_k\|^2] = M\rho\beta_\mathrm{av}$. For i.i.d. Rayleigh fading, \eqref{eq:Eh_hat} becomes
\begin{align}
    \mathbb E[\| \hat{\h}_k\|^2] &= \frac{2\rho}{\pi(\rho K +1)}\tr\big( (\p_k^\tran \Phib^{-1} \p_{k}^*)\otimes \I_M \big) \\
    &= M \frac{2\rho}{\pi(\rho K +1)} \p_k^\tran \Phib^{-1} \p_{k}^* \label{eq:Eh~M}
\end{align} 
with $\C_{\r_{\mathrm{p}}} = (\rho K +1)\Phib\otimes\I_M$ and where the $(u,v)$th element of $\Phib \in \mathbb C^{\tau \times \tau}$ is given by
\begin{align}
\label{eq:Phi} [\Phib]_{u,v} & \triangleq \begin{cases}
1 &\hspace{-21mm} \quad \textrm{if}~u=v, \\
\Omega \Big(\frac{\rho\sum_{k=1}^{K}\Re [P_{u,k}P_{v,k}^*]}{\rho K+1}  \Big) \! - \! j \, \Omega \Big( \frac{\rho\sum_{k=1}^{K}\Im [P_{u,k}P_{v,k}^*]}{\rho K+1} \Big) & \\
& \hspace{-21mm} \quad \textrm{if}~u\neq v.
\end{cases}
\end{align}
Therefore, from \eqref{eq:Eh~M}, $\mathbb E\big[\| \hat{\h}_k\|^2\big]$ increases linearly with $M$. Now, plugging \eqref{eq:h'h (2)} into \eqref{eq:as_orth} yields
\begin{align}\label{eq:h'h (3)}
    & \nonumber\frac{\hat{\h}_k^\herm \hat{\h}_{k'}}{\sqrt{\mathbb E\big[\| \hat{\h}_k\|^2\big] \mathbb E\big[\| \hat{\h}_{k'}\|^2\big]}} \\
    & \to \frac{\rho^2\tr\big( \C_{\h}^{\frac{1}{2}}\bar{\P}^\tran \A_\mathrm{p} \X_k^\herm \X_{k'}\A_\mathrm{p} \bar{\P}^{*}\C_{\h}^{\frac{1}{2}} \big) + \rho\tr\big( \C_{\n}^{\frac{1}{2}}  \X_k^\herm \X_{k'}\C_{\n}^{\frac{1}{2}} \big)}{\sqrt{\mathbb E\big[\| \hat{\h}_k\|^2\big] \mathbb E\big[\| \hat{\h}_{k'}\|^2\big]}} \\& \label{eq:h'h (4)} = \frac{\mathbb E[{\hat{\h}_k^\herm \hat{\h}_{k'}]} }{\sqrt{\mathbb E\big[\| \hat{\h}_k\|^2\big] \mathbb E\big[\| \hat{\h}_{k'}\|^2\big]}}
\end{align}
as $M \to \infty$, where the convergence holds based on \cite[Thm.~3.7]{Couillet-2011}. For i.i.d. Rayleigh fading, \eqref{eq:h'h (4)} becomes
\begin{align}
     \eqref{eq:h'h (4)} &= \frac{\tr(\X_k \C_{\r_{\mathrm{p}}} \X_{k'}^\herm)}{\sqrt{\tr(\X_k \C_{\r_{\mathrm{p}}} \X_k^\herm)\tr(\X_{k'} \C_{\r_{\mathrm{p}}} \X_{k'}^\herm)}} \\& = \label{eq:h'h (5)} \frac{\tr(\bar{\P}_k^\tran \C_{\r_{\mathrm{p}}}^{-1} \bar{\P}_{k'}^*)}{\sqrt{\tr(\bar{\P}_k^\tran \C_{\r_{\mathrm{p}}}^{-1} \bar{\P}_{k}^*) \tr(\bar{\P}_{k'}^\tran \C_{\r_{\mathrm{p}}}^{-1} \bar{\P}_{k'}^*)}} \\& = \label{eq:h'h (6)} \frac{\p_k^\tran \Phib^{-1}\p_{k'}^*}{\sqrt{(\p_k^\tran \Phib^{-1}\p_{k}^*)(\p_{k'}^\tran \Phib^{-1} \p_{k'}^*)}}
\end{align}
where \eqref{eq:h'h (6)} is equal to zero if $\P$ is chosen such that $\P\P^\herm$ is circulant (as in the case of DFT pilot matrix), which implies that $\p_k^\tran \Phib^{-1}\p_{k'}^* = 0$ \cite{Atz22}.
However, for a given $\rho$, we cannot make $\mathbb E[\hat{\h}_k^\herm \hat{\h}_{k'}] \to 0$ and therefore the asymptotic orthogonality does not hold in general. Nevertheless, it can be shown that $\mathbb E[\hat{\h}_k^\herm \hat{\h}_{k'}] \to 0$ as $\tau \to \infty$ and $\rho \to 0$. This is because 
the channel estimation error can be made arbitrarily small by simultaneously increasing $\tau$ and decreasing $\rho$ \cite{Atz22}.

\begin{figure*}[t!]
\setcounter{equation}{80}
\begin{align}
    [\C_{\r|\x}]_{n,m} &=
    \begin{cases}
    \rho K +1 & \quad \textrm{if}~m=n, \\
    \frac{\rho K +1}{2} \Big( \Re \big[ \erf_{\textrm{c}}(\zeta_n) \erf_{\textrm{c}}(\zeta_m^*) \big] +j \, \Im \big[ \erf_{\textrm{c}}(\zeta_n) \erf_{\textrm{c}}(\zeta_m^*) \big] \Big)  & \quad \textrm{otherwise}
    \end{cases}\label{eq:Cr|x (2) (App)}
\end{align}
\hrulefill
\setcounter{equation}{92}
\begin{align}\label{eq:Ek SLS}
    \mathsf{e}_{k}^{\textrm{(MRC)}}(\x,\P) &= M(\rho K+1)\sqrt{\Psi} \sum_{u=1}^{\tau} P_{u,k}^*\bigg[ \Omega\bigg( \frac{\rho\sum_{k=1}^{K} \Re[P_{u,k} x_{k}]}{\sqrt{(\rho K+1)(\rho \sum_{k=1}^{K}|x_{k}|^2 + 1)}} \bigg)  +j \Omega\bigg( \frac{\rho\sum_{k=1}^{K} \Im[P_{u,k} x_{k}]}{\sqrt{(\rho K+1)(\rho \sum_{k=1}^{K}|x_{k}|^2 + 1)}} \bigg) \bigg]
\end{align}
\hrulefill
\vspace{-4mm}
\end{figure*}

\section{Derivations of the LMMD Receiver}\label{sec:App C}

Consider the reformulated receiver design problem in \eqref{eq:V design (2)}, obtained by imposing the MRC receiver in the inner expectation of \eqref{eq:V design (1)}. Then, we use the Bussgang decomposition to rewrite $\r$ in \eqref{eq:r} as
\setcounter{equation}{75}
\begin{align}
  \r &=\G(\x) \y + \d 
\end{align}
where $\d \in \mathbb C^{M}$ is the quantization distortion vector that is uncorrelated with $\y$ and $\G(\x)$ is defined in \eqref{eq:G(x)}. 
Note that, for given $\H$ and $\x$, we have $\y \sim \mathcal{CN}(\sqrt{\rho}\H\x, \I_M)$, which yields
\begin{align} \label{eq:C_{y|x}}
\C_{\y|\x} = \rho \H\x\x^\herm\H^\herm + \I_M.
\end{align}
Furthermore, let $\zeta_m \triangleq \sqrt{\rho}\g_{m}^\tran\x$, where $\g_m \in \mathbb{C}^{M}$ denotes the $m$th column of $\H^\tran$. Then, the $(n,m)$th element of $\C_{\y\r|\x}$ can be written as in \eqref{eq:Cyr|x (1) (App)}--\eqref{eq:Cyr|x (2) (App)} at the top of the page, where the term $\mathbb{E}\big[\Re[y_n]\,\sgn(\Re[y_m])\big]$ is given in \eqref{eq:Ey_n sgn(y_m) (App)} at the top of the page. The remaining terms have a similar form and are thus omitted. Following similar steps, we derive $\C_{\r|\x}\triangleq \mathbb E_{\z}[\r\r^\herm|\x] \in \mathbb{C}^{M\times M}$ as in \eqref{eq:Cr|x (2) (App)} at the top of the next page.
Then, $\C_{\r}$ defined in Section~\ref{sec:Beam} is given by
\setcounter{equation}{81}
\begin{align}
\label{eq:E[rr^H] (App)}
    \C_{\r} = \mathbb E_{\x}[\C_{\r|\x}] = \frac{1}{L^K} \sum_{\x \in \setS^K} \C_{\r|\x}.
\end{align}
Now, the LMMD receiver can be obtained by setting the derivative of the objective of \eqref{eq:V design (2)} to zero, which yields
\begin{align}
\V^{(\mathrm{LMMD})} &= \C_{\r}^{-1} \mathbb{E}_{\x,\z}\big[\r {\e(\x)}^\herm \big]
\end{align}
with $\e(\x)$ defined in \eqref{eq:e(x)}.
Since $\C_{\r}$ can be rank deficient at high SNR, as $\erf(\cdot)$ in \eqref{eq:Cr|x (2) (App)} quickly saturates to $1$ or $-1$, we replace $\C_{\r}^{-1}$ by the pseudoinverse of $\C_{\r}$ and obtain \eqref{eq:WLMMD}.

\section{Formulation of RML} \label{sec:App D}

Let $\tilde{\x} \triangleq \big[ \Re [\x], \Im [\x] \big]^\tran \in \mathbb C^{2K}$, $\tilde{\r} \triangleq \big[ \Re [\r],\Im [\r] \big]^\tran \in \mathbb C^{2M}$, and
\begin{align}
\tilde{\G} \triangleq \begin{bmatrix}
        \Re [\H] & -\Im[\H]\\
        \Im [\H] & \Re[\H]
    \end{bmatrix} ^\tran
    \in \mathbb C^{2K \times 2M}.
\end{align}
RML obtains the data symbol vector $\x^{\star}$ that minimizes the RML function, i.e., \cite{Ngu21}
\begin{align}\label{eq:RML}
    \x^{\star} = \argmin_{\x \in \setS^{K}} \sum_{m=1}^{2M} \mathrm{log} (1 + \mathrm{e}^{ -\theta \tilde{r}_m \tilde{\g}_m ^\tran \tilde{\x}})
\end{align}
where $\tilde{r}_m$ is the $m$th element of $\tilde{\r}$ and $\tilde{\g}_m \in \mathbb C^{2K}$ is the $m$th column of $\tilde{\G}$; furthermore, we have $\theta = 1.702 \sqrt{\frac{4\rho}{\rho K+1}}$.

\section{Analysis of the Probability of Error} \label{sec:App E}

Under i.i.d. Rayleigh fading, the BLMMSE channel estimator in Section~\ref{sec:SM_A} is equivalent to the more tractable scaled least-squares (SLS) channel estimator in \cite{Atz22}.
Using the latter allows to simplify the channel estimate of UE~$k$ as
\begin{align} \label{eq:h_hat SLS}
    \hat{\h}_k &\triangleq \sqrt{\Psi} \R_{\mathrm{p}}\p_k \in \mathbb{C}^M
\end{align} 
with $\Psi$ depending on $\tau$, $\rho$, $K$, and $\P$ \cite{Atz22}. Here, $\R_\mathrm{p} \triangleq Q(\Y_\mathrm{p}) \in \Compl^{M\times\tau}$ denotes the quantized receive signal at the BS with $\Y_\mathrm{p}$ defined in Section~\ref{sec:SM_A}. In the following, we first analyze the distribution of $\xib^{(\textrm{JD})}$ defined in \eqref{eq:xi JD (1)}. Then, we derive the expressions in \eqref{eq:Pr JD} and the
probability of error of UE~$k$ for H-USD. Unless otherwise stated, all the expectations throughout this appendix are taken over $\H$, $\z$, and $\z_{\mathrm{p}}$.

\subsection{Approximate Distribution of $\xib^{(\textrm{JD})}$ in \eqref{eq:xi JD (1)}}\label{sec:App E_A}

Assuming that the MRC receiver is adopted at the BS, the soft-estimated symbol of UE~$k$ is given by $ \hat{x}_{k}^{\textrm{(MRC)}}= \hat{\h}_{k}^\herm \r$. Let $\hat{x}_{\textrm{R},k} \triangleq \Re[\hat{x}_{k}^{\textrm{(MRC)}}]$ and  $\hat{x}_{\textrm{I},k} \triangleq \Im[\hat{x}_{k}^{\textrm{(MRC)}}]$.
Using \eqref{eq:h_hat SLS}, $\hat{x}_{\textrm{R},k}$ can be written as
\setcounter{equation}{86}
\begin{align}
     \hat{x}_{\textrm{R},k} &= \sqrt{\Psi} \Re\big[\p_k^\herm\R_{\mathrm{p}}^\herm \r\big]\label{eq:x_hat_R}
     \\ &= \sqrt{\Psi}\bigg( \Re\bigg[\sum_{m=1}^{M}\Big(\sum_{u = 1}^{\tau}P_{u,k}\big[\R_\mathrm{p}\big]_{m,u}\Big)^*r_m\bigg]\bigg)\label{eq:x_hat_k real sls (1)}
    \\ &= \sum_{m=1}^{M}w_m + \sum_{m=1}^{M}\bar{w}_{m} \label{eq:x_hat_k real sls (2)}
\end{align}
with $w_m \triangleq \sqrt{\Psi}\Re[\phi_m^*r_m] - \bar{w}_{m}$, $\bar{w}_{m} \triangleq \sqrt{\Psi}\mathbb{E}\big[\Re[\phi_m^*r_m]\big]$, and $\phi_m \triangleq \big(\sum_{u = 1}^{\tau}P_{u,k}\big[\R_\mathrm{p}\big]_{m,u}\big)$. Under i.i.d. Rayleigh fading, $\{w_m\}_{m=1}^{M}$ are zero-mean i.i.d. random variables, where each $w_m$ has finite variance $\Gamma_{m}^2 \triangleq \mathbb{E}[|w_m|^2]$. 
Define $\mu_{\textrm{R},k} \triangleq \sum_{m=1}^{M}\bar{w}_{m}$ and $\Sigma_{\textrm{R},k}^2 \triangleq \sum_{m=1}^{M}\Gamma_{m}^2$. Then, the central limit theorem leads to
\begin{align}\label{eq:CLT x_hat_R}
     \lim_{M \to \infty} \hat{x}_{\textrm{R},k} \sim \mathcal{N}(\mu_{\textrm{R},k},\Sigma_{\textrm{R},k}^2).
\end{align}
For any pair $(\bar{\theta}_{1},\bar{\theta}_{2})\in \mathbb{R}^2$, a linear combination between $\hat{x}_{\textrm{R},k}$ and $\hat{x}_{\textrm{R},k'}$, with $k\neq k'$, can be written as
\begin{align}
    \bar{\theta}_{1}\hat{x}_{\textrm{R},k} + \bar{\theta}_{2}\hat{x}_{\textrm{R},k'} &= \sqrt{\Psi}\big(  \bar{\theta}_{1}\Re[\p_k^\herm\R_{\mathrm{p}}^\herm \r] +   \bar{\theta}_{2}\Re[\p_{k'}^\herm\R_{\mathrm{p}}^\herm \r]\big)
    \\& = \sqrt{\Psi}\Re\big[(\bar{\theta}_{1}\p_k+\bar{\theta}_{2}\p_{k'})^\herm\R_{\mathrm{p}}^\herm \r\big]\label{eq:comb x_hat_R}
\end{align}
which is similar to \eqref{eq:x_hat_R}. Since \eqref{eq:CLT x_hat_R} holds for any $\p_k$, then \eqref{eq:comb x_hat_R} approximately follows a Gaussian distribution, and thus the elements of $\Re[\hat{\x}]$ are jointly Gaussian; the same holds for $\Im[\hat{\x}]$. Under i.i.d. Rayleigh fading, the SLS channel estimator allows to simplify the expected values of the soft-estimated symbols in \eqref{eq:Ex MRC} to the form given in \eqref{eq:Ek SLS} at the top of the page. The derivation of \eqref{eq:Ek SLS} follows similar (and more involved) steps as in \cite{Atz22}. Recalling the definition of $\boldsymbol{{\mathsf{e}}}_{\ell_{1},\ldots,\ell_{K}}$ and $\boldsymbol{{\mathsf{e}}}_{t}$ in Sections~\ref{sec:DD_B}~and~\ref{sec:DD_D}, respectively, these vectors comprise the expected values in \eqref{eq:Ek SLS} of the $K$~UEs. Finally, we can rewrite the $t$th element of $\xib^{(\textrm{JD})}$ in \eqref{eq:xi JD (1)} as
 \setcounter{equation}{93}
\begin{align}
[\xib^{(\textrm{JD})}]_t &=-2\Re\big[\hat{\x}^\herm \f_t(\x)\big] + \|\boldsymbol{{\mathsf{e}}}_{t}\|^2-\|\boldsymbol{{\mathsf{e}}}_{\ell_{1},\ldots,\ell_{K}}\|^2\label{eq:xi JD (2)}
\end{align}
with $t \in \mathcal{T}$, and where $\mathcal{T}$ and $\f_t(\x)$ are defined in Section~\ref{sec:DD_B}. Since $\xib^{(\textrm{JD})}$ is an affine function of $\Re[\hat{\x}]$ and $\Im[\hat{\x}]$, for large $M$, it approximately follows $\mathcal{N}\big(\mub^{(\textrm{JD})}(\x),\boldsymbol{\Sigma}^{(\textrm{JD})}(\x)\big)$ with $\mub^{(\textrm{JD})}(\x)$ and $\boldsymbol{\Sigma}^{(\textrm{JD})}(\x)$ defined in \eqref{eq:mu xi}--\eqref{eq:Sigma xi}.

\subsection{Upper Bound on the Probability of Error for JD (Derivation of \eqref{eq:Pr JD})}\label{sec:App E_B}

Let $\mathcal{L}$ denote the set of all possible group indices $l_1,\ldots,l_K$ for $l_k = \ell_{k}$, with $|\mathcal{L}| = L^{K-1}$. Recalling the data detection method presented in Section~\ref{sec:DD_B} as JD, the exact probability of error of UE~$k$ for a given $\x$ is given by
\begin{align}
\nonumber \mathsf{P}_{k}^{\textrm{(JD})}(\x) = \mathbb{P}\Big[ & \minn_{t \in \mathcal{T}} \|\hat{\x} - \boldsymbol{{\mathsf{e}}}_{t} \|^2 \\
\label{eq:SER JD analysis 1} & \leq \minn_{(l_1,\cdots, l_K) \in \mathcal{L}} \|\hat{\x} - \boldsymbol{{\mathsf{e}}}_{l_1,\ldots,l_K} \|^2 \Big|\x\Big] 
\end{align}
where the right-hand side of the inequality in \eqref{eq:SER JD analysis 1} is no greater than $\|\hat{\x} - \boldsymbol{{\mathsf{e}}}_{\ell_{1},\ldots,\ell_{K}} \|^2$.
Hence, an upper bound on \eqref{eq:SER JD analysis 1} is given by 
\begin{align}
\eqref{eq:SER JD analysis 1} &\leq \mathbb{P}\Big[\minn_{ t \in \mathcal{T}} \|\hat{\x} - \boldsymbol{{\mathsf{e}}}_{t} \|^2 \leq  \|\hat{\x} - \boldsymbol{{\mathsf{e}}}_{\ell_{1},\ldots,\ell_{K}} \|^2\,\Big|\x\Big]
\\& = \mathbb{P}\Big[\minn_{ t \in \mathcal{T}} \big(\|\hat{\x} - \boldsymbol{{\mathsf{e}}}_{t} \|^2 -\|\hat{\x} - \boldsymbol{{\mathsf{e}}}_{\ell_{1},\ldots,\ell_{K}} \|^2 \big)\leq 0\,\Big|\x\Big]\\&=\overline{\mathsf{P}}_{k}^{\textrm{(JD})}(\x) \label{eq:SER JD analysis 2}.
\end{align}
Recalling $\xib^{\textrm{(JD)}}$ from \eqref{eq:xi JD (1)}, we can rewrite \eqref{eq:SER JD analysis 2} as
\begin{align}
\overline{\mathsf{P}}_{k}^{\textrm{(JD})}(\x) &= \Pr\big[\minn_{t \in \mathcal{T}}\, [\xib^{(\textrm{JD})}]_t \leq 0\big|\x\big]\label{eq:SER analysis 1}
   \\& = 1 - \Pr\big[\minn_{t \in \mathcal{T}} \,[\xib^{(\textrm{JD})}]_t \geq 0\big|\x\big]\label{eq:SER analysis 2}
   \\&= 1 - \Pr\big[\xib^{(\textrm{JD})} \geq \0_{(L-1)L^{K-1}}\big|\x\big].\label{eq:SER analysis 3}
\end{align} 
Finally, using the approximate distribution of $\xib^{\textrm{(JD)}}$ in Appendix~\ref{sec:App E_A}, the expression in \eqref{eq:SER analysis 3} can be derived as in \eqref{eq:Pr JD}.

\subsection{Probability of Error for H-USD}\label{sec:App E_C}

Recalling the data detection method presented in Section~\ref{sec:DD_A} as H-USD (equivalent to $N$-JD with $N=1$), for a given $\x$, the exact probability of error of UE~$k$ is given by
\begin{align}\label{eq:SER analysis H-USD}
\mathsf{P}_{k}^{\textrm{(H-USD})}(\x) & = \mathbb{P}\big[\!\minn_{l_k\in \{1,\cdots, L\}/ \ell_{K}}\!\!\! |\hat{x}_{k} - \bar{\mathsf{e}}_{k,l_{k}}|^2 \leq  |\hat{x}_{k} - \bar{\mathsf{e}}_{k,\ell_{k}}|^2 \big|\x\big]  
\end{align}
with $\bar{\mathsf{e}}_{k,l_{k}}$ defined  in \eqref{eq:E_kl}. Following the same steps as in \eqref{eq:SER JD analysis 2}–\eqref{eq:SER analysis 3}, $\mathsf{P}_{k}^{\textrm{(H-USD})}(\x)$ can be obtained in a form analogous to \eqref{eq:Pr JD}, with a definition of the detection error vector similar to \eqref{eq:xi JD (1)}.

\addcontentsline{toc}{chapter}{References}
\bibliographystyle{IEEEtran}
\bibliography{refsabbr,refs}

\end{document}